\documentclass[twocolumn,tighten]{aastex631}
\hypersetup{linkcolor=red,citecolor=blue,filecolor=cyan,urlcolor=magenta}

\usepackage{apjfonts,amsfonts,amsmath,amssymb,bm,verbatim,float}

\definecolor{RedWine}{rgb}{0.743,0,0}
\definecolor{GrassGreen}{rgb}{0.125,0.75,0.125}
\definecolor{RoyalBlue}{rgb}{0.25,0.41,0.88}
\definecolor{darkgreen}{cmyk}{0.85,0.2,1.00,0.2} 
\definecolor{purple}{cmyk}{0.5,1.0,0,0} 
\definecolor{ultramarine}{rgb}{0.07, 0.04, 0.56}
\definecolor{cadmiumgreen}{rgb}{0.0, 0.42, 0.24}
\definecolor{indigo(dye)}{rgb}{0.0, 0.25, 0.42}

\usepackage[caption=false]{subfig}

\begin{document}

\title{
Confronting the Diversity Problem: The Limits of Galaxy Rotation Curves as a tool to Understand Dark Matter Profiles
}
\shorttitle{Confronting the Diversity Problem}

\shortauthors{Sands et al.}
\correspondingauthor{Isabel Sands}
\email{isands@caltech.edu}
\author[0000-0002-1159-4882]{Isabel S.~Sands}
\affiliation{TAPIR, Mailcode 350-17, California Institute of Technology, Pasadena, CA 91125, USA.}
\author[0000-0003-3729-1684]{Philip F.~Hopkins}
\affiliation{TAPIR, Mailcode 350-17, California Institute of Technology, Pasadena, CA 91125, USA.}
\author[/0000-0002-6196-823X]{Xuejian Shen}
\affiliation{Kavli Institute for Astrophysics and Space Research, Massachusetts Institute of Technology, Cambridge, MA, 02139, USA}
\author[0000-0002-9604-343X]{Michael Boylan-Kolchin}
\affiliation{Department of Astronomy, The University of Texas at Austin, 2515 Speedway, Stop C1400, Austin, Texas 78712-1205, USA}
\author[0000-0003-4298-5082
]{James Bullock}
\affiliation{Department of Physics and Astronomy, University of California Irvine, Irvine, CA 92697, USA}
\author[0000-0002-4900-6628]{Claude-André Faucher-Giguère}
\affiliation{Department of Physics and Astronomy and CIERA, Northwestern University, Evanston, IL 60208, USA}
\author[0000-0002-5908-737X]{Francisco J.~Mercado}
\affiliation{Department of Physics and Astronomy, Pomona College, Claremont, CA 91711, USA}
\author[0000-0002-3430-3232]{Jorge Moreno}
\affiliation{Department of Physics and Astronomy, Pomona College, Claremont, CA 91711, USA}
\affiliation{Center for Computational Astrophysics, Flatiron Institute 162 Fifth Avenue, New York, NY 10010, USA}
\author[0000-0003-2806-1414]{Lina Necib}
\affiliation{Kavli Institute for Astrophysics and Space Research, Massachusetts Institute of Technology, Cambridge, MA, 02139, USA}
\author[0000-0002-4669-9967]{Xiaowei Ou}
\affiliation{Kavli Institute for Astrophysics and Space Research, Massachusetts Institute of Technology, Cambridge, MA, 02139, USA}
\author{Sarah Wellons}
\affiliation{Department of Astronomy, Van Vleck Observatory, Wesleyan University, 96 Foss Hill Drive, Middletown, CT 06459, USA}
\author[0000-0003-0603-8942]{Andrew Wetzel}
\affiliation{Department of Physics \& Astronomy, University of California, Davis, CA, USA 95616}

\begin{abstract}

While galaxy rotation curves provide one of the most powerful methods for measuring dark matter profiles in the inner regions of rotation-supported galaxies, at the dwarf scale there are factors that can complicate this analysis. Given the expectation of a universal profile in dark matter-only simulations, the diversity of observed rotation curves has become an often-discussed issue in Lambda Cold Dark Matter cosmology on galactic scales. We analyze a suite of Feedback in Realistic Environments (FIRE) simulations of $10^{10}-10^{12}$ $M_\odot$ halos with standard cold dark matter, and compare the true circular velocity to rotation curve reconstructions. We find that, for galaxies with well-ordered gaseous disks, the measured rotation curve may deviate from true circular velocity by at most $\sim$ 10\% within the radius of the disk. However, non-equilibrium behavior, non-circular motions, and non-thermal and non-kinetic stresses may cause much larger discrepancies of $\sim$ 50\% or more. Most rotation curve reconstructions underestimate the true circular velocity, while some reconstructions transiently over-estimate it in the central few kiloparsecs due to dynamical phenomena. We further demonstrate that the features that contribute to these failures are not always visibly obvious in HI observations. If such dwarf galaxies are included in galaxy catalogs, they may give rise to the appearance of ``artificial" rotation curve diversity that does not reflect the true variation in underlying dark matter profiles. 
\end{abstract}

\keywords{galaxy: evolution ---
          galaxy: rotation curves ---
          dark matter ---
          dwarf galaxies ---
          methods: numerical}

\section{Introduction}

Measurements of galaxy rotation curves (RCs) are one of the most effective methods of probing underlying dark matter halo density profiles \citep{1978ApJ...225L.107R,1978PhDT.......195B}. By comparing these inferred halo density profiles to numerical simulations, it has been determined that cold dark matter-only simulations over-predict the density in the innermost regions of the halo, while observations indicate the presence of a flattened ``core" in the halo center \citep{1994Natur.370..629M,2008AJ....136.2761O,2015AJ....149..180O}. Various resolutions to this ``cusp-core problem" have been proposed, such as the critical role that baryonic feedback plays in shaping the potential in the central region of the halo (see \cite{delpopolo2022review} and references therein). Other proposed solutions to the cusp-core problem involve modifications to the dark matter itself, such as the addition of self-interactions \citep{Tulin_2018}.

As a result, gas rotation curves for dwarf galaxies can serve as a probe of potential deviations from Lambda Cold Dark Matter ($\Lambda$CDM) cosmology on small scales \citep{Weinberg_2015}. At present, these rotation curve measurements are the only detailed probe of dark matter physics at the kiloparsec scale. Additionally, beyond the original core-cusp problem, other small-scale tensions between observations and simulations with CDM have arisen, such as the ``too big to fail" and ``plane of satellites" problems \citep{Boylan_Kolchin_2011,Wetzel_2016,Bullock_2017}. 

Recent surveys of dwarf galaxies as tests of small-scale structure and alternative dark matter have revealed a greater diversity of rotation curves than predicted by numerical simulations \citep{Oman_2015}. This so-called ``diversity problem" extends beyond the traditional cusp-core problem: observations have provided evidence for the existence of highly-concentrated dwarf galaxies, as well as extended, less-dense cores. So far, numerical hydrodynamic simulations with CDM have not been able to reproduce the observed diversity of rotation curves \citep{Applebaum_2021, 2016MNRAS.nihao, 2016MNRAS.apostle, Garrison_Kimmel_2019}. Notably, it is the fastest-rising rotation curves (corresponding to the most compact dwarf galaxies) that are the most difficult to reproduce in hydrodynamical simulations. 

Of all of the small-scale tensions in cosmology, the diversity problem arguably poses the most direct challenge to $\Lambda$CDM \citep{2022NatAs...6..897S}. While some of the spread in observed rotation curves can likely be explained by baryonic feedback \citep{El_Badry_2017}, the difficulty of reproducing the most extreme measurements in hydrodynamic simulations with CDM has served as a compelling motivation for studies of alternative models of dark matter. For instance, the formation of concentrated dwarfs may be facilitated by gravothermal runaway and collapse in self-interacting dark matter (SIDM) models \citep{Zentner_2022,Vargya_2022, Jiang_2023}. Several studies have found that SIDM models are able to replicate the diversity of observed rotation curves in the Spitzer Photometry and Accurate Rotation Curves (SPARC) catalog \citep{Lelli_2016}: while some work has shown no statistical preference between SIDM and CDM models with baryonic feedback \citep{Zentner_2022}, other work has suggested that SIDM may be a better fit to the SPARC dataset \citep{PhysRevX.9.031020, Kaplinghat_2020}. 

However, all of these studies assume that the fitted rotation curves of observed galaxies accurately trace the underlying dark matter distribution and thus give the correct enclosed mass of the galaxy. Measured rotation curves are an inferred quantity that make fundamental assumptions about the system they probe: that the galaxy is in steady-state equilibrium and is spherically symmetric. Moreover, rotation curves assume that full phase space information about the gas and stars can be recovered from three degrees of freedom (radial position, position along axis of rotation, and rotational velocity) out of seven total (6D phase space and time). As new studies probe rotation curves at higher precision and pose questions about more subtle features such as the shapes of rotation curves, it is questionable whether these assumptions still apply. For instance, \cite{Downing_2023} argue that a large fraction of rotation curves could be affected by complex non-circular motions that are not captured by modeling. Additionally, it has been demonstrated in other astrophysical systems that an incorrect assumption of steady-state equilibrium can lead to errors in inferred quantities: stellar dynamical Jeans modeling can fail to reconstruct the enclosed mass by up to a factor of 2 due to the effects of non-equilibrium dynamics \citep{El_Badry_2017_jeans}. Moreover, previous work has shown that turbulent pressure, non-circular motions, and out-of-equilibrium effects can cause significant deviations in the inferred rotation curves of simulated galaxies at high redshifts ($z \gtrsim 1$) \citep{2020MNRAS.497.4051W}.

As the resolution of surveys continues to improve, so does our ability to ask more detailed questions about the behavior and structure of dark matter in dwarf galaxies. While order-of-magnitude consistency between observations and simulations once sufficed, improved observations and simulations can now render meaningful, more subtle differences between data and $\Lambda$CDM predictions at these scales, with the important caveat that the comparison must be made in an apples-to-apples manner. In this work, we investigate the limits of how well rotation curves can probe underlying dark matter profiles at the dwarf galaxy scale. We focus on a sample of FIRE dwarf galaxies mostly within the stellar mass range of $10^8-10^9 M_{\odot}$, where baryonic feedback and bursty star formation can result in highly non-equilibrium structure. This is also the mass scale at which deviations from $\Lambda$CDM cosmology have been most studied, and the scale at which other studies of dwarf galaxies from cosmological zoom-in simulations have shown that non-equilibrium effects are most important \citep{2014MNRAS.437..415D, Chan_2015, 2015MNRAS.454.2092O}.

This paper is organized as follows: $\S$\ref{section:methods} briefly summarizes the methods used for the simulations and their analysis; $\S$\ref{section:results} provides an overview of the FIRE-3 dwarf galaxies analyzed and their resulting rotation curve behavior; $\S$\ref{section:obs} compares the simulations to observations of dwarf galaxies that exemplify the diversity problem; $\S$\ref{section:conclusion} summarizes our findings and primary conclusions.

\section{Methods}
\label{section:methods}

\subsection{FIRE simulations}

In this work, we analyze simulations from the Feedback in Realistic Environments (FIRE) project \footnote{\url{http://fire.northwestern.edu}}. Most simulations were run with the ``FIRE-3" version of the code, the details of which are described in section 2 of \citet{Hopkins_2022}. These simulations were run with the same set of parameters, physics, and source code as the simulations detailed in \cite{Hopkins_2022}. The FIRE simulations use the GIZMO code \citep{Hopkins2015gizmo}\footnote{\url{http://www.tapir.caltech.edu/~phopkins/Site/GIZMO.html}}, a massively-parallel, multi-method code with multiple fluid equation solvers. We set spatial resolution for hydrodynamics and gravitational force-softening with a fully-adaptive Lagrangian method at fixed mass resolution. The hydrodynamic equations are solved with the Lagrangian Godunov meshless finite-mass (MFM) method. 

In these simulations, the criteria for star formation requires that gas is molecular, dense, self-gravitating, self-shielding, and Jeans-unstable \citep{2014MNRAS.445..581H,Hopkins_2022}. Stellar feedback occurs due to AGB and OB mass loss from stellar winds, type Ia and II supernovae, and multi-wavelength photo-heating and radiation pressure. Cooling and heating in the simulations are driven by the metagalactic background and stellar sources between $10-10^{10}$K.

In this paper, we consider a sample of fourteen low-to-intermediate mass FIRE-3 dwarf galaxies, plus one Milky Way-mass galaxy used as a benchmark for comparison. We analyze only the most massive, primary central galaxy within each simulation volume except when two galaxies of similar mass are in the process of merging. We select the center of each galaxy by identifying the point of highest stellar mass concentration; however, in some cases (e.g. mergers), this method may fail to correctly select the kinematic and dynamical center of the system, in which case we identify the center as the median position of star particles. We discuss the phenomena that complicate the selection of the galactic center as well as the impact on inferred rotation curves in sections $\S$\ref{sec:mergers} and $\S$\ref{sec:feedback}. Snapshots are taken at 10 Myr intervals for all halos except for m11a, in which the interval is 1 Myr. We also consider one simulation, ``m11e-2", which was run with the ``FIRE-2" version of the code \citep{ El_Badry_2017, 2018MNRAS.480..800H}; we choose to include this simulation in our analysis in order to have an additional example of a galaxy merger to study tidal effects on rotation curves.

\subsection{Rotation Curves}

\begin{table*}
    \hspace*{-1cm} 
    \centering
    \begin{tabular}{|c|c|c|c|c|c|c|}
    \hline
         \bf{Reconstruction} & $\frac{\partial(\rho\bf v)}{\partial t}  - \bf{S}_{\rm{ext}}$ & $\boldsymbol{\Pi}^{\rm{non-kin}}_{\rm{non-therm}}$ & $\boldsymbol{\Pi}^{\rm{kin}}_{\rm{i,j}}$ & $\boldsymbol{\Pi}^{\rm{kin}}_{\rm{non-circ}}$   & $\boldsymbol{\Pi}_{\rm{therm}}$ & $\langle \boldsymbol{\Pi}^{\rm{kin}}_{\rm{\phi \phi}} \rangle$\\ \hline
         Full RC & \checkmark  & \checkmark & \checkmark  & & \checkmark & \\ \hline
         Equilibrium &  & \checkmark & \checkmark &  &\checkmark & \\ \hline 
         Non-circular Recovery &  &  &  & \checkmark & \checkmark & \checkmark \\ \hline
         Coherent Centrifugal &  & & & & & \checkmark \\ \hline
    \end{tabular}
    \caption{A summary of the terms included in each rotation curve reconstruction used in this work. The first column, $\frac{\partial (\rho {\bf v} )}{\partial t} - \bf{S}_{\rm{ext}}$, denotes the time-dependent, non-equilibrium term and external source terms from radiation pressure and cosmic rays. The second column is defined as the non-thermal, non-kinetic components of the stress tensor: $\boldsymbol{\Pi}^{\rm{non-kin}}_{\rm{non-thermal}} \equiv \boldsymbol{\Pi}_{\rm{mag}} + \boldsymbol{\Pi}_{\rm{visc}} + \boldsymbol{\Pi}_{\rm{cr}} $. The third column is the full kinetic stress tensor, $\boldsymbol{\Pi}^{\rm{kin}}_{ij}$, where $i,j$ are indices corresponding to the $r, \theta, \phi$ components. The fourth column refers to only the diagonal elements of the kinetic stress tensor, $\boldsymbol{\Pi}^{\rm{kin}}_{\rm{non-circ}} \equiv \boldsymbol{\Pi}^{\rm{kin}}_{rr} + \boldsymbol{\Pi}^{\rm{kin}}_{\theta \theta} + \boldsymbol{\Pi}^{\rm{kin}}_{\phi \phi}$ (with corrections made to account for isotropic and line-of-sight velocity dispersions).  The fifth column denotes thermal pressure. Finally, the last column denotes the centrifugal component of the kinetic stress tensor: $\langle \boldsymbol{\Pi}^{\rm{kin}}_{\phi \phi} \rangle = \langle \rho \bf v_{\phi} \bf v_{\phi}\rangle$.}
    \label{tab:rc_reconstructions}
\end{table*}

In simulations, by definition, we numerically integrate the momentum equation, defined as 
\begin{align}
\frac{\partial (\rho {\bf v} )}{\partial t} + \nabla \cdot \boldsymbol{\Pi}^{\ast} = - \rho\nabla \Phi + {\bf S}_{\rm ext},
\end{align}
where $\Phi$ is gravitational potential, and ${\bf S}_{\rm ext}$ is the external source term due to radiation and cosmic rays. We define ${\bf S}_{\rm ext} \equiv \rho\, {\bf a}_{\rm rad} + {\bf s}_{\rm cr} $, where $\rho$ is the matter density, ${\bf a}_{\rm rad} \equiv  \frac{1}{c}\int\,\kappa_{\nu}{\bf F}_{\nu}\,d\nu $ is the acceleration due to radiation pressure, and ${\bf s}_{\rm cr} \equiv -\frac{1}{c^{2}} \hat{\bf B}\, D_{t} F_{\rm cr,\,e}$ is the out-of-flux-equilibrium cosmic ray residual. In these definitions, $\bf F_i$ denotes the flux of a species $i$, $D_{t}$ is the conservative comoving derivative, and $\hat{\bf B}$ is the unit magnetic vector field. 
Both ${\bf a}_{\rm rad}$ and $ {\bf s}_{\rm cr}$ are negligibly small in the simulations. $\boldsymbol{\Pi}^{\ast}$ is the stress tensor, which includes components from the physics modeled in the FIRE-3 simulations. We broadly categorize these components as kinetic stress, stress due to isotropic thermal pressure, and non-thermal stress due to magnetic fields, viscosity of the ISM, and cosmic rays: 
\begin{align}
\boldsymbol{\Pi}^{\ast} &\equiv \boldsymbol{\Pi}_{\rm kin} + \boldsymbol{\Pi}_{\rm mag} + \boldsymbol{\Pi}_{\rm therm} + \boldsymbol{\Pi}_{\rm visc} + \boldsymbol{\Pi}_{\rm cr},
\label{eqn:pi_comps}
\end{align}
with full mathematical expressions for each component given in Appendix \ref{section:app_RC}. 

We can rearrange the momentum equation to see the explicit relation to acceleration due to gravity: 
\begin{align}
-\nabla \Phi &\equiv {\bf a}_{\rm grav} \equiv \frac{1}{\rho} \left[  \frac{\partial (\rho {\bf v} )}{\partial t} + \nabla \cdot \boldsymbol{\Pi}^{\ast} - {\bf S}_{\rm ext} \right].
\label{eqn:momentum}
\end{align}
With several simplifying assumptions to equation \ref{eqn:momentum}, we can recover the relation between circular velocity and enclosed mass for an idealized system, in which the gravitational force and centrifugal force within a galaxy are perfectly balanced. First, we assume that the system is in perfect steady-state equilibrium, so the time derivative vanishes. We further assume that the gravitational potential is perfectly spherical, and that the kinetic stress tensor is completely dominated by uniform, homogeneous circular motion in the galactic plane. In this case, the only non-vanishing term of the stress tensor is $\Pi^{*}_{\phi \phi}$. Finally, we assume that the external source terms due to radiation pressure and cosmic rays are negligible. When all of these assumptions hold, the vertical and tangential components of equation \ref{eqn:momentum} vanish, and the radial component gives equation \ref{eqn:vc}, the familiar equation for acceleration under a centrifugal force:

\begin{align}
\langle \langle a_{\rm grav} \rangle \rangle (r) \equiv \frac{V_{\rm c}^{2}}{r} = \frac{G\,M_{\rm enc}(<r)}{r^{2}}
\label{eqn:vc}
\end{align}
where $V_c$ is the circular velocity, and $M_{\rm enc}$ is the total enclosed mass at radius $r$. The double angular brackets denote that we take both an average over annuli and time.

As we drop these assumptions and allow a more general framework, the acceleration due to gravity, $\bf{a}_{\rm{grav}}$, along with the $V_c$ derived from it, will begin to deviate from the simple proportionality to enclosed mass in equation \ref{eqn:vc}. The primary exercise of this work is to determine how relaxing these assumptions in turn affects the inferred gravitational acceleration in simulations of dwarf galaxies. To this end, we can understand in which limits each approximation may fail. In order to create the rotation curve reconstructions, we output additional information at late-time FIRE simulation snapshots, including acceleration due to gravity, hydrodynamic acceleration, HI gas velocity and density gradients, radiation pressure and flux, and all components of the stress tensor (kinetic, thermal, magnetic fields pressure, and Braginsky viscosity).  

We consider several approximations to the fully general equation \ref{eqn:momentum} in order to compose our rotation curve reconstruction schemes. First, we assume all components of the stress tensor, non-equilibrium terms, and external source terms are measurable, and average over the gravitational acceleration in an annulus in the plane perpendicular to the $z$-component of the angular momentum vector. We refer to this as the ``full rotation curve reconstruction", and it accurately reproduces the true $V_c$ curve as defined in equation \ref{eqn:vc} as long as the system is spherically symmetric. Then, we assume the galaxy is in steady-state equilibrium, and drop the time-derivative term in equation \ref{eqn:momentum}. We further drop the external source term, which is usually negligible compared to the components of the stress tensor. The reconstructed RC with these terms is referred to as the ``equilibrium rotation curve reconstruction". Next, we consider the components of the stress tensor that cannot be measured from observations. These include the tensor components pertaining to the magnetic field, viscosity, and cosmic rays, which are often neglected in observational RC analysis. We refer to this limit in the text as the ``non-circular recovery reconstruction". Finally, we make the most stringent assumption: that the only measurable component of velocity is the centrifugal $v_{\phi}$, and that the line-of-sight (LOS) dispersion components from thermal pressure cannot be recovered; this is the ``coherent centrifugal reconstruction". 

While we have also considered combinations of terms that are not shown in this paper, these four rotation curve reconstructions encapsulate our key conclusions about the qualitative ways in which rotation curves may fail to recover the true $V_c$. We summarize the constituent terms of each RC reconstruction in Table \ref{tab:rc_reconstructions} for convenience and clarity. In this work, we use neutral atomic hydrogen (HI) in order to reconstruct the rotation curves for each galaxy; the neutral hydrogen fraction of each gas element is calculated by GIZMO, and we exclude gas identified as molecular \citep{ORR2018MNRAS.478.3653O}. While observations of other atomic and molecular spectral lines can be used to measure rotation curves, we limit this study only to the simulated HI due to its abundance in spiral galaxies, where the HI often extends beyond the visible disk \citep{Sofue_2001}. Each radial bin contains at minimum $\mathcal{O}(100)$ particles in the inner few points, and $\mathcal{O}(1000)$ particles at larger radii, making the RC reconstructions robust to sampling.

\begin{table*}
\centering
\begin{tabular}{|c|c|c|c|c|c|c|}
\hline
     Name & $r_{\rm{vir}}$ (kpc) & $M_{\rm{halo}}$ ($M_{\odot}$)& $M_{\ast}$  ($M_{\odot}$)& $M_{\rm{HI}}$ ($M_{\odot}$) & HI disk morphology & RC success \\ \hline
     \multicolumn{7}{|c|}{\bf{Classical Dwarfs}} \\
     \hline 
      m10q & 230 & $7.9\times 10^{9}$ & $9.1\times 10^{6}$ & $5.9\times 10^{3}$ & Insufficient HI & N/A \\ \hline
      m10v & 230 &  $8.3 \times 10^{9}$ & $4.1 \times 10^{6} $& $17$ & Insufficient HI & N/A  \\ \hline
      m10b & 230 & $9.2 \times 10^{9}$ &  $1.0\times 10^{7}$ & $2.9 \times 10^{5}$ & Small, puffy disk & Deviates due to thermal pressure \\ \hline
      \multicolumn{7}{|c|}{\bf{Bright Dwarfs}} \\ \hline
      m11a & 230 & $2.9 \times 10^{10}$ & $3.5 \times 10^{8}$ & $2.7 \times 10^{6}$ & Non-disky & Deviates due to non-circular motions \\ \hline
      m11b & 230 & $3.3 \times 10^{10}$ & $6.1 \times 10^{8}$ & $7.9 \times 10^{6}$ & Small, puffy disk & Accurate within disk \\ \hline
      m11v & 230 & $ 4.6 \times 10^{10}$ & $8.2 \times 10^{8}$ & $2.3 \times 10^{7}$ & Small, puffy disk & Accurate within disk \\ \hline 
      m11i & 240 & $ 5.0 \times 10^{10}$ & $2.8 \times 10^{9}$ & $2.1 \times 10^{8}$ & Turbulent disk & Deviates due to bursty star formation\\ \hline 
      m11c &  230 & $8.1 \times 10^{10}$ & $5.1 \times 10^{8}$ & $6.4 \times 10^{6}$ & Non-disky & Deviates due to strong B-fields \\ \hline
      m11e & 240 & $ 8.4 \times 10^{10}$ & $3.0 \times 10^{9}$ & $1.0 \times 10^{8}$ & Disk w/ tidal features & Deviates due to merger \\ \hline
      m11q & 230 & $ 8.5 \times 10^{10}$ & $2.0 \times 10^{9}$ & $1.2 \times 10^{5}$ & Small, puffy disk & Deviates due to non-circular motions\\ \hline
      m11e-2 \footnote{m11e-2 is run with FIRE-2 physics; all other simulations are run with FIRE-3.} & 270 & $8.7\times 10^{10}$ & $1.6 \times 10^{9}$ & $1.0 \times 10^{9}$ & Disk w/ tidal features & Deviates due to merger \\ \hline
      m11h & 270 & $ 1.0 \times 10^{11}$ & $6.1 \times 10^{9}$ & $7.9 \times 10^{6}$ & Small, puffy disk & Accurate within disk \\ \hline 
      m11d & 240 & $ 1.1 \times 10^{11}$ & $4.7 \times 10^{9}$ & $3.8 \times 10^{5}$ & Ordered disk & Accurate to within $\sim$ 10\%\\ \hline
      m11f & 270 & $ 2.1 \times 10^{11}$ & $5.9 \times 10^{9}$ & $1.3 \times 10^{8}$ & Ordered disk & Accurate to within $\sim$ 10\% \\ \hline 
      m11g & 300  & $ 2.2 \times 10^{11}$ & $1.5 \times 10^{10}$ & $2.0 \times 10^{8}$ & Ordered disk & Accurate to within $\sim$ 10\%\\ \hline
      \multicolumn{7}{|c|}{\bf{Milky Way Analog}} \\ \hline
      m12i & 370 & $ 4.6 \times 10^{11}$ & $ 3.6 \times 10^{10}$ & $ 3.3 \times 10^{9}$ & Ordered disk & Accurate to within $\sim$ 10\% \\\hline 
\end{tabular}
\caption{A summary of the simulated FIRE galaxies analyzed for this work, along with their estimated virial radii (in kpc); halo, stellar, and HI masses within $\frac{1}{5}r_{\rm{vir}}$ of the galactic center (in $M_{\odot}$), a brief description of their morphology, and a description of the behavior of the RC reconstructions. Within the sample of fifteen galaxies, there is a wide range of HI disk morphologies and dynamical behaviors.}
\label{table:galaxies}
\end{table*}

\section{Simulation Results}
\label{section:results}

\begin{figure*}
    \centering
    \gridline{\fig{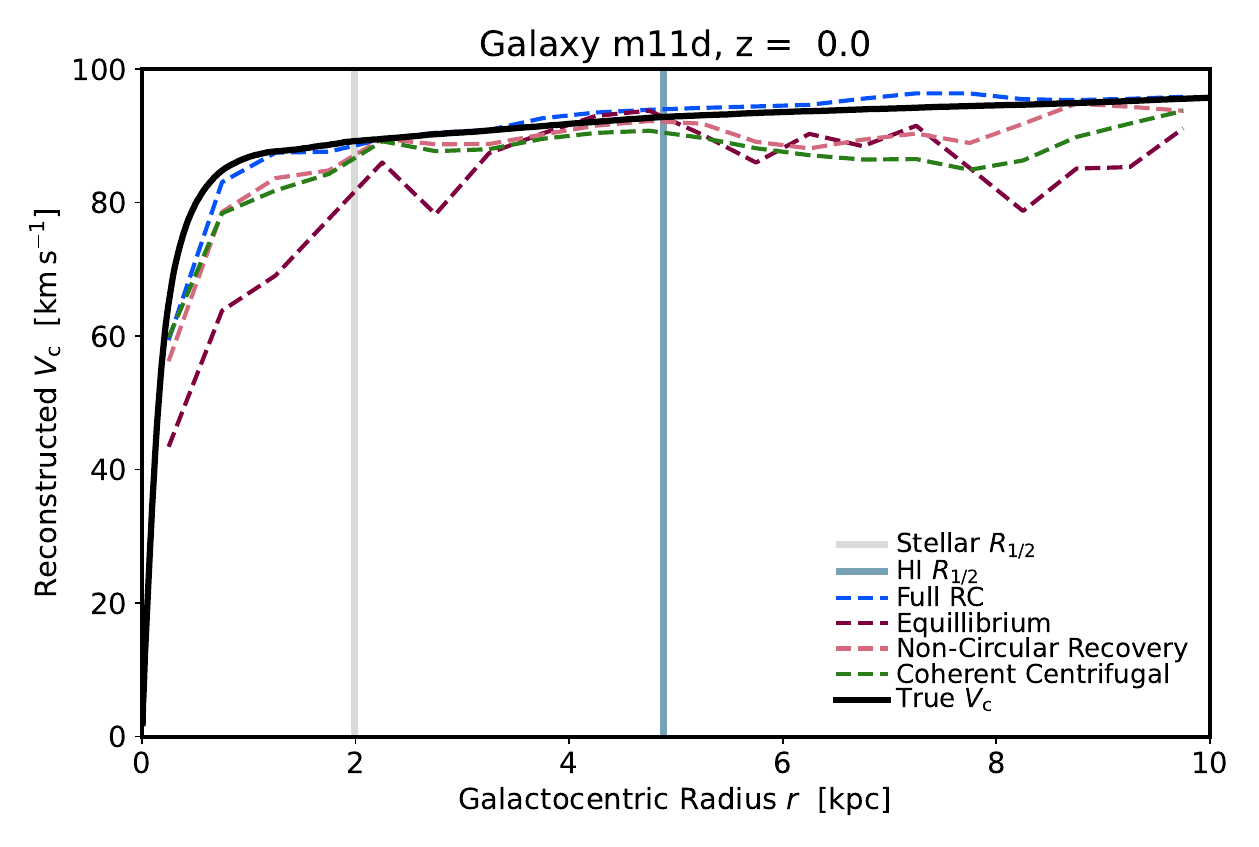}{0.49\textwidth}{(a) Ordered disk} \fig{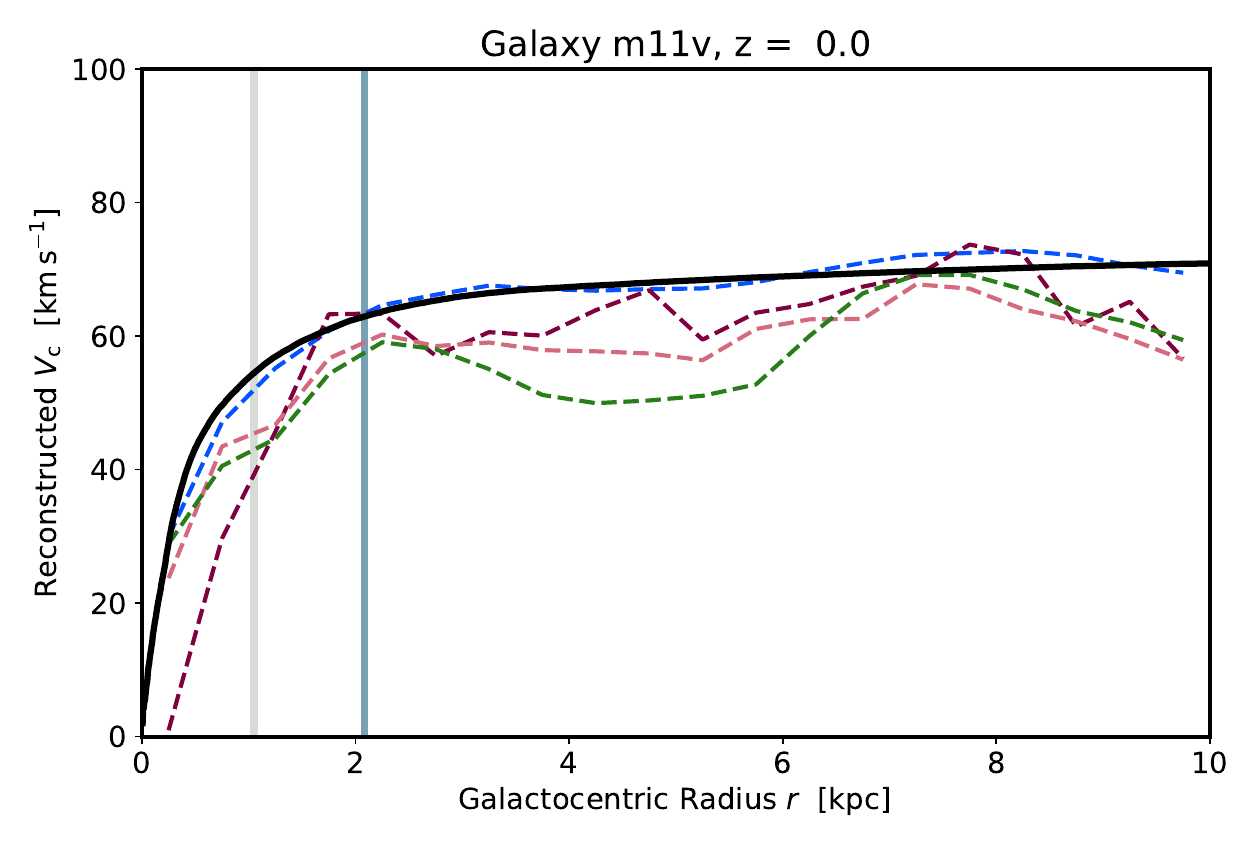}{0.49\textwidth}{(b) Small, puffy disk}}
    \gridline{\fig{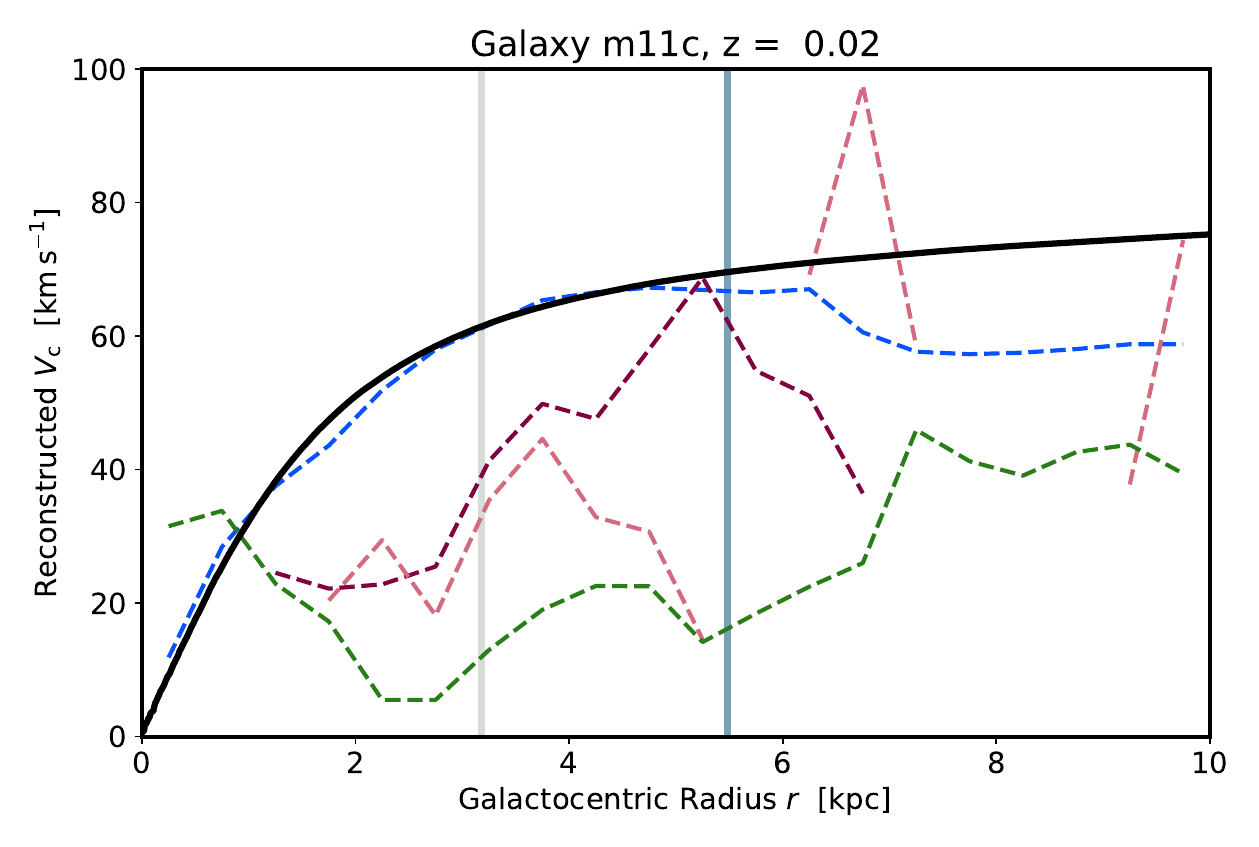}{0.49\textwidth}{(c) Strong magnetic field, no disk} \fig{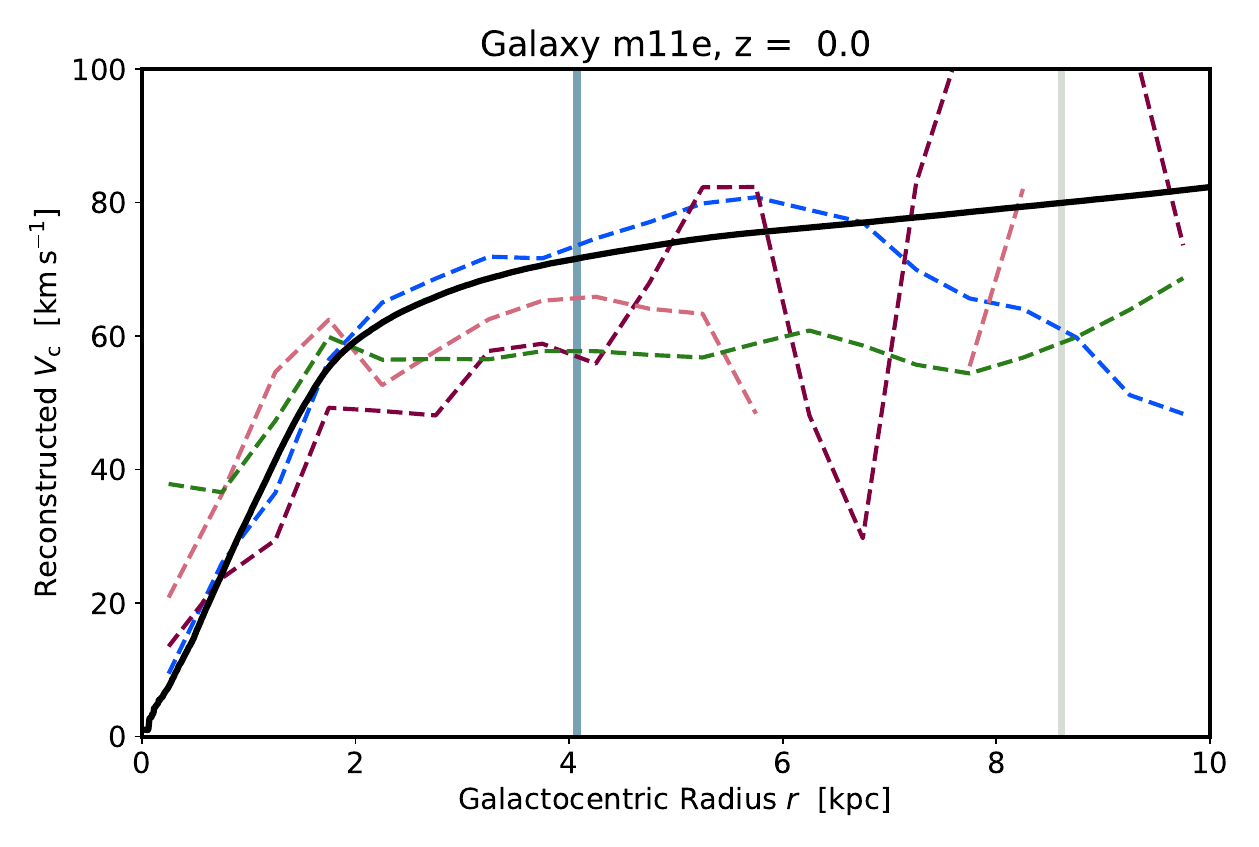}{0.49\textwidth}{(d) Ongoing interaction between two galaxies}}

    \caption{Reconstructed rotation curves vs true circular velocity for (clockwise, starting at top left) galaxies m11d, m11v, m11e, and m11c. The true circular velocity defined in equation \ref{eqn:vc} , which maps the enclosed mass at radius $r$, is plotted in solid black. Four rotation curve reconstructions are shown. The blue curve is the ``perfect" reconstruction, which contains all terms in equation \ref{eqn:momentum}, weighted in an annular average. The red curve drops the time derivative term, the pink curve drops the external source term and all non-kinematic, non-thermal stress tensor terms. Finally, the green curve plots the circular velocity depending only on coherent centrifugal motion. These four galaxies broadly encapsulate the classes of behavior we find in our sample. m11d has a well-ordered thin disk and, accordingly, fairly accurate RC reconstructions. m11v has a small, puffy disk, and the RC reconstructions map the true $V_c$ well in the inner few kpc, but features apparent in the reconstructions beyond the disk do not reflect the true mass distribution.  m11e is a system undergoing a merger, and the RC reconstructions for the larger of the two galaxies deviate from the true $V_c$ due to tidal effects and out-of-equilibrium dynamics. Note that the HI half-mass radius appears closer to the galactic center than the stellar half-mass radius for m11e at this redshift; this is a reflection of the dynamics of the merger. Finally, m11c has strong magnetic fields and, as a result, the RC reconstructions fail when the non-equilibrium term and relevant stress tensor component are dropped.}
    \label{fig:RC_grid_60}
\end{figure*}

In this section, we describe the results for the rotation curve analysis from our sample of FIRE-3 galaxies. Some of the RC reconstructions are quite accurate; others have order unity failures for myriad reasons, which we shall discuss more at length in the following subsections. In Figure \ref{fig:RC_grid_60}, we show the rotation curve reconstructions for a sub-sample of four FIRE-3 galaxies selected to demonstrate the modes of RC reconstruction success and failure. The galaxies shown in the top row of Figure \ref{fig:RC_grid_60}, m11d and m11v, both have fairly accurate RC reconstructions within the HI half-mass radius (light blue), while the galaxies in the bottom row, m11c and m11e, have reconstructions that deviate significantly from the true $V_c$

Three m11 (bright dwarf) galaxies  in our sample reproduce the true $V_c$ to within approximately 10\% accuracy out to the HI half-mass radius. As an example, Figure \ref{fig:RC_grid_60} (upper left) shows several reconstructed rotation curves for one such FIRE-3 galaxy, m11d, compared to the true $V_c$. These galaxies generally feature the most ordered, flat, extended disks, with coherent centrifugal motion of HI in the disk. These features are evident from the HI maps (Figure \ref{fig:m11d_HI}). The RC reconstructions for m12i are also accurate; m12i is the standard FIRE Milky Way analog, has a very thin disk, and is largely in steady state equilibrium, so this behavior is expected \citep{2016ApJ...827L..23W}.

In all other galaxies in our sample, the reconstructed rotation curves depart more significantly from the true $V_c$. In some m11 dwarfs, the accuracy of rotation curve reconstruction depends on the amount of HI present. Often, there is sufficient HI in the inner portions of the galaxy, resulting in accurate reconstructions in this region. However, beyond the scale of a couple of kpc, the disk ends and the HI density falls off. At this point, the rotation curve reconstructions may deviate significantly from the true $V_c$ (Figure \ref{fig:RC_grid_60}, top right). These disks tend to be smaller and relatively ``puffy" when compared to the disky galaxies described above-- that is, the height-to-radius ratio is larger. The HI disk in all four of these galaxies extends to a radius of 2-4 kpc, with less dense regions of HI for several kpc beyond. The RC reconstructions for these galaxies should serve as a caution against attempting to extrapolate RCs beyond the HI half-mass radius. 

In other cases, the galaxy may lack a disk altogether, or rotation may not be the dominant kinetic term in the stress tensor. These galaxies can be identified by their HI density plots, and by evaluating the norms of each component of equation \ref{eqn:pi_comps}. Two galaxies in our sample fall into this category: m11a, which has significant bulk radial motion, and m11c, which is subject to strong magnetic fields. m11a and m11c are relatively small and ``puffy," with a more spherical or ellipsoid HI distribution than a true disk (Figure \ref{fig:RC_grid_60}, bottom left). Galaxy m11a is relatively spherical, with significant inflows, to which the bulk radial motions can be attributed. m11c has less coherent structure, and little evidence of a disk.

Many of the largest deviations from true $V_c$ in the rotation curve reconstructions occur when non-equilibrium effects become important. We see one such highly non-equilibrium system in m11e, in which two dwarf galaxies merge with a resulting halo mass of approximately $10^{11} M_{\odot}$.  During the merger, we see that there are large jumps in the reconstructed RCs, as well as significant information lost when dropping non-equilibrium terms (Figure \ref{fig:RC_grid_60}, bottom right). This result is consistent with the consensus from observations and simulations that galaxy mergers are out-of-equilibrium systems \citep{Hung_2015, Barrera_2015, bloom_2017, McElroy_2022}. Another galaxy in our sample, m11i, has significant non-equilibrium effects due to bursty star formation and stellar feedback at late times. 

We include three m10 (classical dwarf) galaxies in our sample.  One of these galaxies (m10q) is lacking in HI, and is far below the threshold of detectability even by state-of-the-art instruments. A second galaxy, m10v, has enough HI to be feasibly detectable, but it is concentrated in several molecular clouds at any given snapshot; there is no disk structure or rotation to allow accurate RC analysis. The third m10 galaxy, m10b, shows rotational motion, and has a small, ``puffy" disk.

The galaxies in our sample can be categorized by how well their RC reconstructions reproduce the true circular velocity, and the reasons why the RC reconstructions may fail. In this section, we evaluate the cases of both minor deviations from true $V_c$, as well as more catastrophic failures. In the latter case, the RC reconstructions for these galaxies most often fail due to insufficient HI in the disk or elsewhere in the galaxy. These HI-deficient galaxies are less of a concern, as measurements of their observed analogs should indicate the extent of the HI disk, or they may be too faint to observe in the first place. However, there are several more interesting cases where simulated galaxies with inaccurate RC reconstructions are less obviously problematic. 
\vspace{0.5cm}

\subsection{Galaxies where Rotation Curve Analysis Fits Well}

The primary criteria that determine the success of rotation curve reconstructions is whether the galaxy forms a well-ordered disk with a sufficient amount of HI. The four galaxies in our sample (m11d, m11f, m11g, and m12i) whose RC reconstructions most faithfully recover the true $V_c$ all have ordered, thin, extended disks. It is not particularly surprising that such galaxies produce fairly accurate RC reconstructions: thin, disky galaxies that are dominated by circular motion are among the closest to the systems for which equation \ref{eqn:vc} can be expected to hold. This class of bright dwarfs has more in common with Milky Way analogs like m12i than their less massive counterparts; of the bright dwarfs analyzed in this sample, they are the most similar to the spiral galaxies from benchmark RC studies, such as \cite{de_Blok_2008_things}. 

These three galaxies are the most massive bright dwarfs in the sample, in terms of both stellar and halo mass (Table \ref{table:galaxies}). The amount of HI gas present seems to matter less in determining the success of the RC reconstructions, as long as there is a sufficiently detectable amount of HI (column density $ N_{\rm{HI}} \gtrsim 10^{-18} \rm{HI cm}^{-2}$); while m11d is approximately half of the halo mass of m11f and m11g, it has approximately a thousandth of their HI mass out to one-fifth of the virial radius. That more massive halos are more likely to form well-ordered disks is not surprising: previous simulations have found that the factors that can quantify diskiness tend to be a function of mass \citep{Hopkins_2023}. 

\begin{figure*}
    \centering      
    \includegraphics[width=\textwidth]{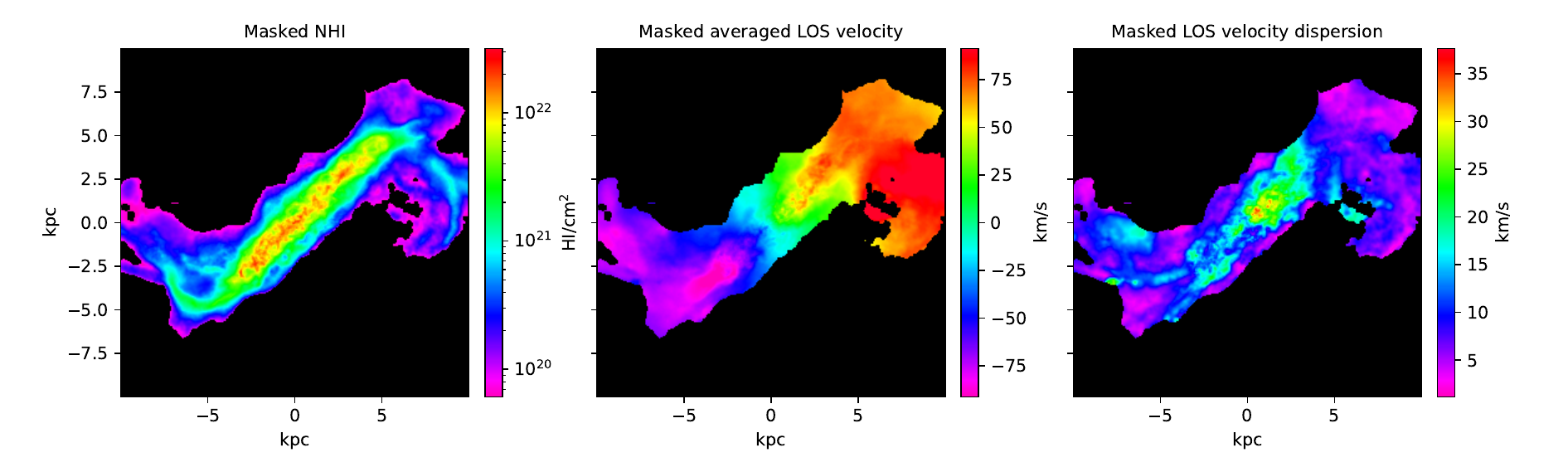}
    \caption{HI column density, first kinematic moment, and second kinematic moment for galaxy m11d at $z = 0$, which is an example of a galaxy where the RC analysis fits well; these plots are ``masked" such that only regions with HI column density $ N_{\rm{HI}} \gtrsim 5 \times 10^{-19} \rm{HI cm}^{-2}$ are shown. When HI column density is viewed approximately edge-on, the most dense region lies within the plane of the disk, with spiral arms visible at the sides. The average LOS-velocity has a clear gradient across the center of the galaxy. The LOS velocity dispersion is fairly symmetric, with the highest velocity dispersion at the center of the galaxy. The presence of an ordered disk, velocity gradient, and symmetry seen in the HI distribution indicate that this galaxy is a good candidate for RC analysis, and that the RC should map the underlying mass distribution relatively faithfully.}
    \label{fig:m11d_HI}
\end{figure*}

The visual markers of disk formation and spiral structure are obvious in HI density and velocity maps for these galaxies. Figure \ref{fig:m11d_HI} shows these features for galaxy m11d: the HI disk appears thin when viewed edge-on, and the spiral arms are visible when viewed face-on or at a slight inclination angle. The average LOS velocity shows a clear gradient about the center of the galaxy, indicating that the disk is dominated by rotational motion, and the LOS velocity dispersion is symmetric about the center of the galaxy, where it peaks. 

The small deviations from true $V_c$ for these galaxies are due to many of the same reasons that more extreme failures in RC reconstruction can occur, namely non-equilibrium behavior due to bursty star formation, as well as bulk radial inflows and outflows. However, in a galaxy with a well-ordered disk, this behavior occurs to a lesser degree, and has less overall impact on galaxy morphology and on the kinematics of gas and stars \citep{Hopkins_2023}. In general, the minor deviations from the true $V_c$ in such galaxies tend to be underpredictions of the circular velocity, especially in the central few kiloparsecs. Previous studies have attributed this under-estimate to a net inflow of gas through the disk; since this is a non-equilibrium effect, it cannot be captured by observations, and the information loss occurs from dropping the time derivative term in equation \ref{eqn:momentum} \citep{Hopkins_2019}. It is important to note that even a discrepancy at the 10\% level may not be negligible in the discussion of small scale features in rotation curves, such as ``bumps" after a sharp rise in the inner few kpc.

One notable feature within this class of galaxies (visible in Figure \ref{fig:RC_grid_60}, top left) is that, often, the coherent centrifugal reconstruction (green curve) more accurately recovers the true $V_c$ than the equilibrium reconstruction (red curve), despite the latter including more terms in the stress tensor. This behavior is due to the fact that certain terms in the stress tensor, most notably non-circular motion, arise from non-equilibrium phenomena. When the time-dependent term is dropped, but the resulting non-zero kinematic components of the stress tensor remain in equation \ref{eqn:momentum}, then the equation is unbalanced, and the equilibrium reconstruction falls below the other rotation curve reconstructions. When the corresponding term in the stress tensor is also dropped in the coherent centrifugal reconstruction, this effect cancels out.

\subsection{RC reconstruction failure beyond the extent of the HI disk}

Many of the bright dwarfs in our sample form small, puffy HI disks; while they are supported by rotation, they also tend to be more turbulent and disrupted than the larger, more ordered disks formed by the more massive m11 galaxies described above. These disks are smaller in radius and have a higher height-to-radius ratio than the galaxies described in the previous section. These galaxies are not necessarily deficient in HI overall: m11b, m11h, and m11v span a total HI mass of approximately $10^5-10^7 M_{\odot}$, which is comparable to galaxies like m11d for which we are able to more accurately recover the true circular velocity. Nor are these galaxies especially low in overall halo mass relative to others in the sample. 

Within this class of dwarf galaxies, there is enough HI to create faithful RC reconstructions up to the extent of the disk (roughly 2-4 kpc). Beyond the radius of the disk, however, the RC reconstructions deviate significantly from the true circular velocity. In this work, we use the HI half-mass radius to estimate the extent of the HI disk in our sample of simulated dwarfs. Beyond this radius, non-equilibrium effects may also plays a role, especially in the  circumgalactic medium (CGM) of dwarf galaxies. For the bright dwarfs, the inner CGM is primarily cool and supported by turbulence-like inflows onto and outflows from the disk rather than rotation \citep{2017ARA&A..55..389T, 2023ARA&A..61..131F}. All of these factors contribute to the difficulty of recovering the true circular velocity beyond the extent of the HI disk. This result is not particularly surprising, but serves as a caution against extrapolating about extending rotation curve analysis to galaxies with very low HI surface brightness.   

Two of the classical dwarfs in our sample (m10q and m10v), have very little HI throughout their halos, and do not form any kind of disk-like structure. The third, m10b, has more total HI and does form a small, puffy quasi-disky structure that is partially supported by rotation.  However, m10b (as well as m10v) is quenched, and the dominant stress tensor term is the thermal pressure from the CGM. As a result of the thermal pressure dominating over kinematic rotation in the m10s, the RC reconstructions for these galaxies fail \citep{Wheeler_2016}.

Observed analogs to these galaxies in our sample would be easily identified as poor candidates for RC analysis: the m10s would likely be too faint to be detected. HI observations of size and morphology of the m11s would indicate the regions in which RC analysis can be trusted. These measures could potentially lead to a biased catalog, but would not introduce artificial rotation curve diversity.

\subsection{Catastrophic Failures within the HI disk}
\label{sec:cat_fail}

We now turn to the cases where a galaxy's reconstructed RCs may differ significantly from the true $V_c$ by a factor of up to order unity. It is difficult to quantify a concrete failure rate because the FIRE-3 simulations analyzed in this work are isolated halos, and are not drawn from a representative volume. Instead, we focus on several galaxies in our sample that provide interesting case studies of systems where RC reconstructions fail due to complex galactic dynamics or the effects of baryonic feedback. Within our overall sample, almost every galaxy is subject to some amount of non-equilibrium behavior that can cause the reconstructed RCs to deviate from the true circular velocity. In most cases, these deviations are minor and not the primary cause of a catastrophic failure; however, within our sample we find evidence of galaxies where non-equilibrium physics leads to deviations of order unity. 

These galaxies pose a fundamental problem for observations: they have sufficient HI for detection, but the phenomena responsible for deviations in their reconstructed RCs cannot be observed. Any non-equilibrium phenomena is inherently undetectable because observing the galaxy at one time fails to capture information about the time derivative term in equation \ref{eqn:momentum}. Many galaxies within this category also exhibit significant non-circular motions. While modeling can correct for eccentricities in the motion of the gas, these corrections are limited by the loss of two of the three velocity dimensions when observing the distribution of HI. Additionally, the tilted ring modeling used to construct rotation curves from observations relies upon prior assumptions about the motion of gas in a galaxy (e.g. concentric ellipses) \citep{1989A&A...223...47B}. As shown in this section, there is not a simple or obvious prior that can capture the true HI distribution or kinematics for these outliers. 

The behaviors leading to large deviations from the true circular velocity in RC reconstructions are, by far, most common in dwarf galaxies in the stellar mass range of $10^7-10^9 M_{\odot}$. This mass scale is the where bursty star formation peaks, giving rise to a peak in complex galactic dynamics \citep{Garrison_Kimmel_2019}. In galaxies above this mass scale, star formation is less messy and less involved in inducing the formation of cores; below this mass scale, bursty star formation is less dynamically important. This dwarf galaxy mass range is also important to answering questions about the observed distribution and behavior of dark matter (e.g. cold vs self-interacting). Thus, poor RC fits from galaxies in this mass range can lead to incorrect conclusions about dark matter in the regime of most interest for testing it. 

\subsubsection{Failure due to Gravitational Dynamics}
\label{sec:mergers}

In galaxy m11e, two dwarf galaxies merge at late time and form a disk. Throughout the merger and even after the disk in the resulting galaxy has begun to settle, there are significant non-equilibrium effects. In addition, there is a torque on the disk, so the velocity cross-terms in equation \ref{eqn:kin_therm} are non-negligible. These terms contribute to the catastrophic failure of RC reconstructions for this galaxy. RC analysis of m11e is further complicated by the difficulty of choosing a galactic center in a system where the center-of-mass is rapidly evolving, or has no obvious placement. Typically, we assume that the stars, gas, and dark matter in a galaxy share the same kinematic and dynamical center. However, in a system that is out-of-equilibrium, or with two interacting structures of similar mass, this is not necessarily the case. Ambiguity in selecting the center is inevitable in such systems, and can contribute to inaccuracies in recovering the true $V_c$.

\subsubsection{Failure due to Feedback}
\label{sec:feedback}

\begin{figure*}
    \centering
    \includegraphics[width=\textwidth]{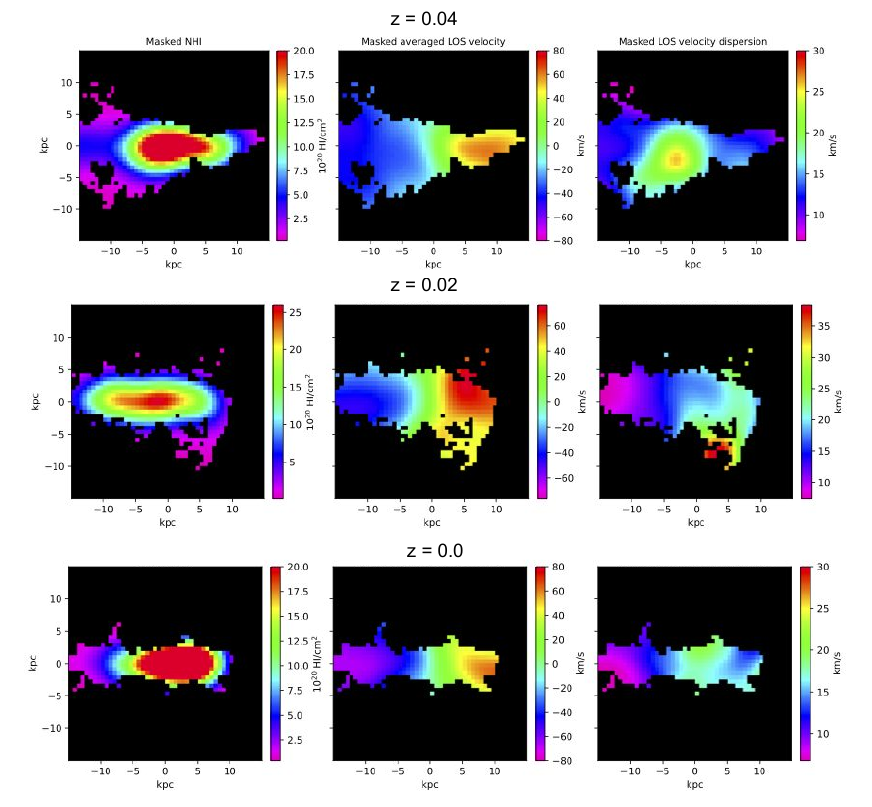}
    \caption{ HI maps for FIRE galaxy m11i, which is an example of a turbulent disk with bursty star formation that varies in structure over time, at an edge-on view for three late-time snapshots ($z=0.04$, $z=0.02$, and $z=0.0$). From left to right, HI column density, average LOS velocity, and LOS velocity dispersion are shown. The maps have been pixelated and a Gaussian blur applied in order to compare to HI maps for observed dwarfs.}
    \label{fig:m11i_HI}
\end{figure*}

The magnitude of non-equilibrium effects on a single galaxy's dynamics may vary significantly with time. This phenomenon is exemplified by m11i, where non-equilibrium terms lead to highly inaccurate RC reconstructions in some late-time snapshots. These deviations are largely attributable to a burst of star formation at late times, with the star formation rate peaking at approximately $0.4$ Gyr in lookback time. 

\begin{figure}
    \centering
    \includegraphics[width=0.5\textwidth]{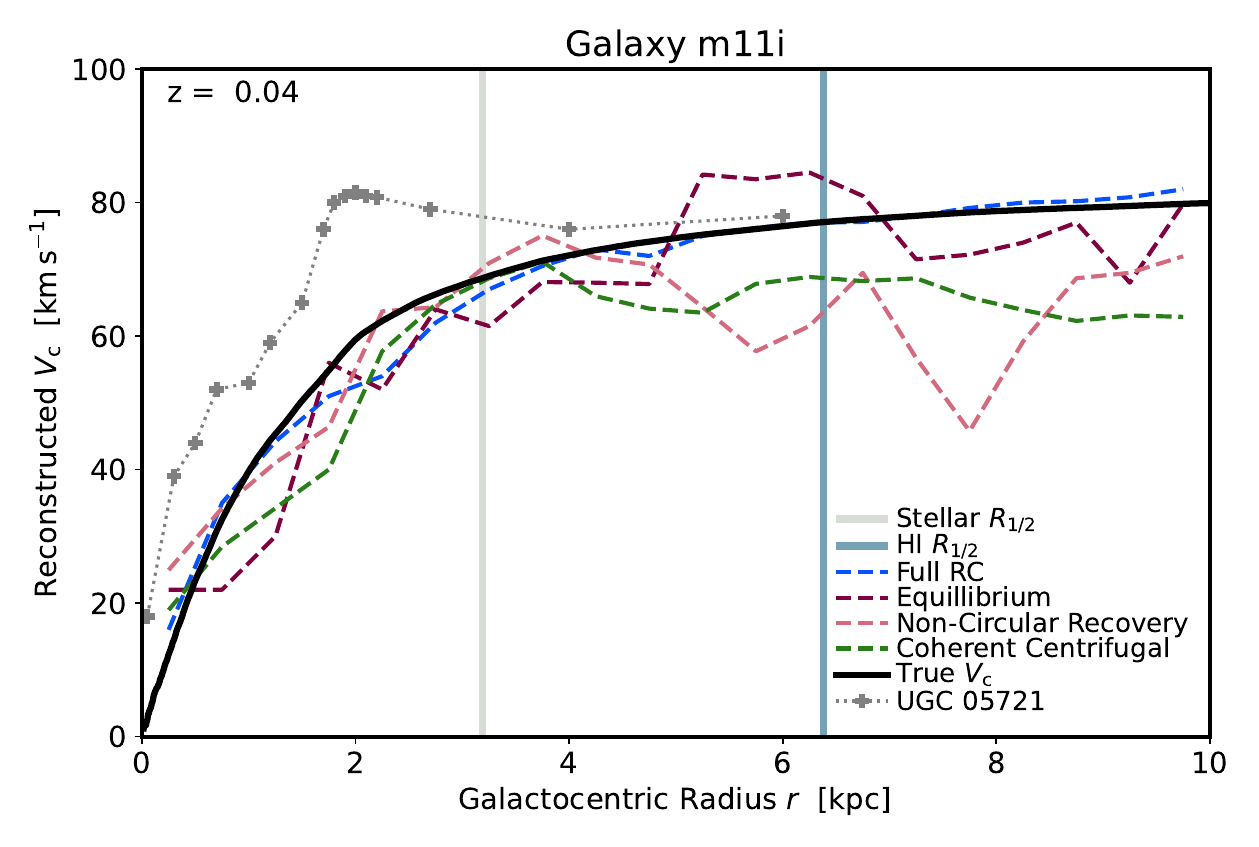}
    \includegraphics[width=0.5\textwidth]{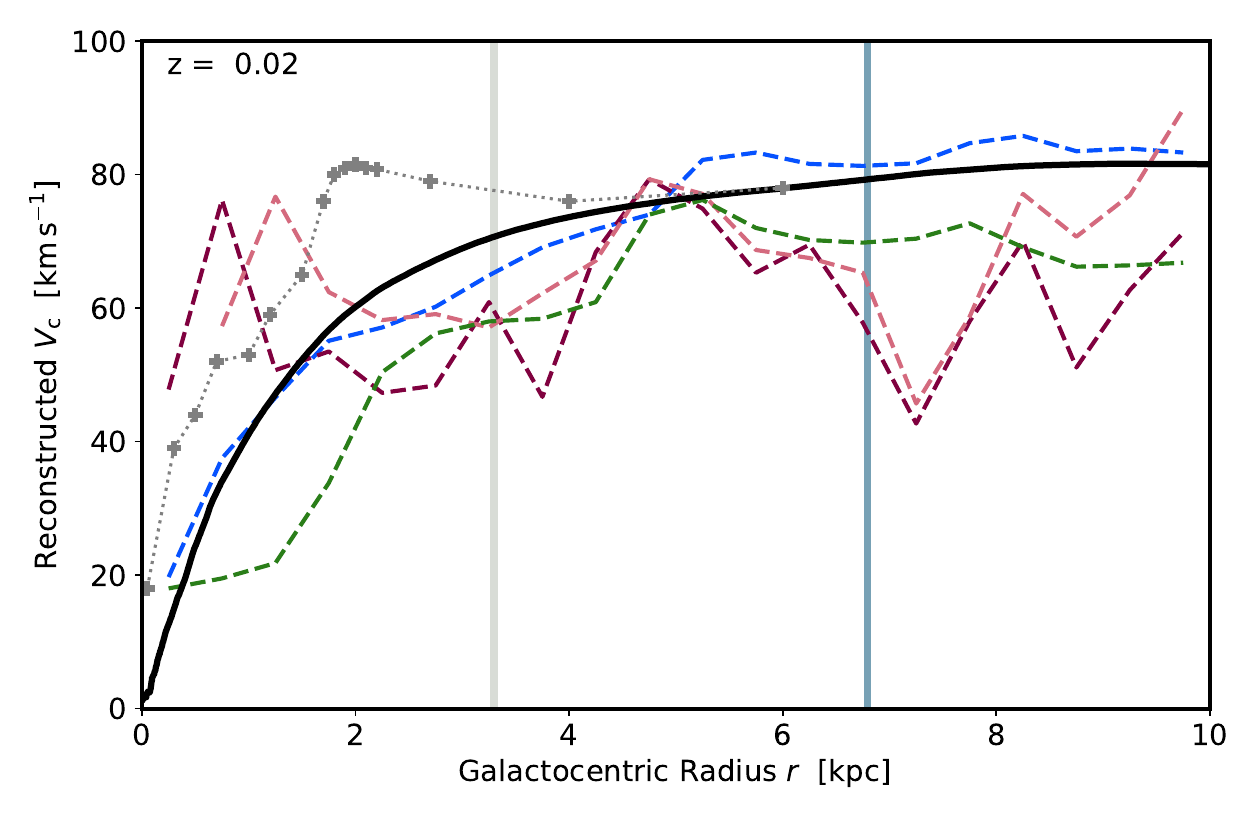}
    \includegraphics[width=0.5\textwidth]{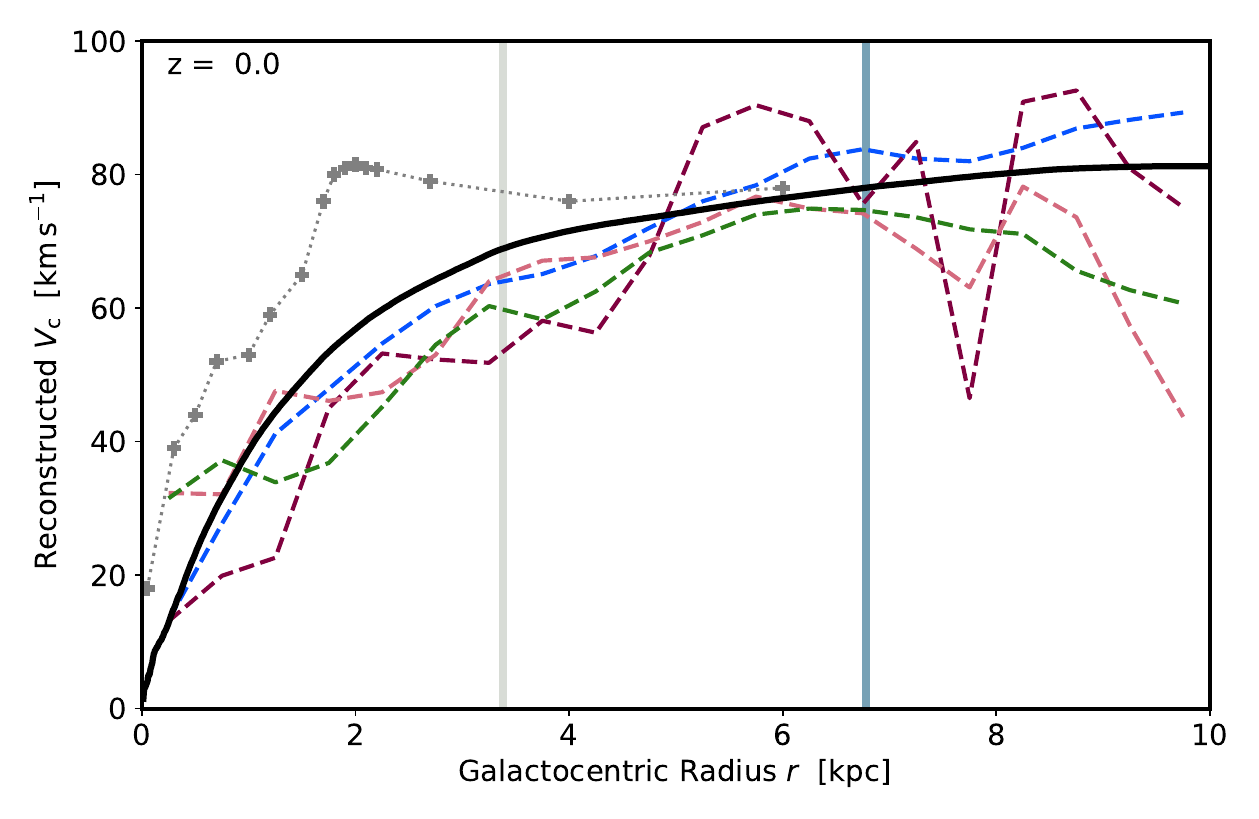}
    \caption{Reconstructed RCs for galaxy m11i at the same three late-time redshifts shown in Figure \ref{fig:m11i_HI}. The reconstructions are compared to the observed RC for UGC 5721, often cited in the literature as an example of a compact dwarf that exemplifies one extreme of the diversity problem \citep{Oman_2015}. At $z = 0.02$, some RC reconstructions for m11i shows a similar peak to UGC 5721, but this vanishes at earlier and later times due to transient, non-equilibrium phenomena.}
    \label{fig:m11i_rc}
\end{figure}

While galaxy m11i does form a HI disk, it is consistently disrupted by this bursty star formation. As shown in Figure \ref{fig:m11i_HI}, the primary effect of the late-time peak in star formation is to introduce asymmetries in the disk, as well as outflows of gas, which can vary significantly between snapshots. These features are highly non-equilibrium phenomena; as a result, the rotation curve reconstructions at these snapshots also vary in accuracy. Because the non-equilibrium term cannot be captured by observations, one could draw very different conclusions when looking at rotation curves from a highly dynamical galaxy like m11i based solely on the time at which it was observed. This result is demonstrated at three different late-time snapshots in  Figure \ref{fig:m11i_rc}. The efficacy of the RC reconstructions is best for the inner few kpc at $z = 0.04$ and $z = 0$, where the HI density maps show that the inner portion of the disk is relatively thin and ordered, but the outer portions are more disrupted. 

Figure \ref{fig:m11i_HI_angles} shows m11i at $z = 0.02$ from several different viewing angles. Although m11i at $z = 0.02$ shows a thinner HI disk than the other two snapshots, its LOS velocity gradient and velocity dispersion still reveal a significant degree of asymmetry. When viewed face-on, the source of the asymmetries becomes clear: there is a hole in the HI distribution that arises from a late burst of star formation. As a result, the stellar, gas, and dark matter centroids do not coincide in this galaxy. This lack of symmetry and the offset between gas, stars, and dark matter complicate the selection of the galactic center, causing the equilibrium and non-circular recovery RC reconstructions to spike in the inner two kpc. We note that the peaks in these RC reconstructions are suggestive of observed RCs corresponding to supposedly compact dwarf galaxies, such as UGC 5721, the rotation curve for which is plotted in gray. We will discuss this result in the context of the RC diversity problem in section \ref{section:div_prob}.

\subsubsection{Failure due to complex non-circular motion}

In some cases, non-equilibrium behavior may also feed into terms in the stress tensor, leading other sources of stress and pressure to dominate over rotation. In galaxy m11a, bulk radial outflows and inflows dominate the stress tensor, and the resulting RC reconstructions depending on centrifugal motion fail. In some regions of this galaxy, the effect of thermal pressure becomes non-negligible. All of these factors are inherently linked to non-equilibrium effects,  thus the RC reconstructions that drop time-dependent terms also fail to recover the true circular velocity. Galaxy m11q shows similar features, but the magnitude of the deviations from true $V_c$ is smaller. 

\subsubsection{Failure due to non-thermal stress}

While less common in our sample than non-equilibrium behavior, there are rare cases where non-kinetic terms in the stress tensor lead to failures in RC reconstructions. We find one example of a galaxy, m11c, where the dominant stress tensor component is from the magnetic field contribution. In this system, the magnetic field pressure can counteract the gravitational force, which leads to significant under-predictions of the circular velocity of HI gas in RC reconstructions where the magnetic field component of the stress tensor is dropped. Strong magnetic fields in a galaxy can also make the system more dynamic, leading to the large fluctuations in the reconstructed RCs where the non-equilibrium terms are excluded \citep{Dobbs_2007}. 

\section{Connection to Observations}
\label{section:obs}

\subsection{Contamination by poor RC analysis candidates}

Within a relatively small sample of simulated galaxies, we are able to find a number of cases where RC analysis fails due to complex galactic dynamics in disky galaxies, which begs the question: how often can we expect such galaxies to be mistaken as better candidates for RC analysis than they are in actuality? The most important criteria for predicting the efficacy of RC reconstructions is the diskiness of the galaxy; however, a galaxy can appear disky and still have compounding factors that result in inaccurate RC reconstructions.

High-quality observational samples of dwarf galaxies with rotation curves select targets to include in their analyses in order to minimize the errors described in the previous sections \citep{de_Blok_2008_things, 2012AJ....144..134H}. Nonetheless, for dwarf galaxies within the mass range of greatest interest for tests of dark matter, ideal disky galaxies are quite rare. Pragmatically, the dwarfs selected for rotation curve analysis have a height-to-radius ratio of less than 0.5 \citep{ScaleHeight2020MNRAS.495.2867P}, and while a velocity shear is usually visible, there are often substantial asymmetries in the velocity dispersion \citep{HI_dwarf}. 

\begin{figure}
    \centering
    \includegraphics[width=0.5\textwidth]{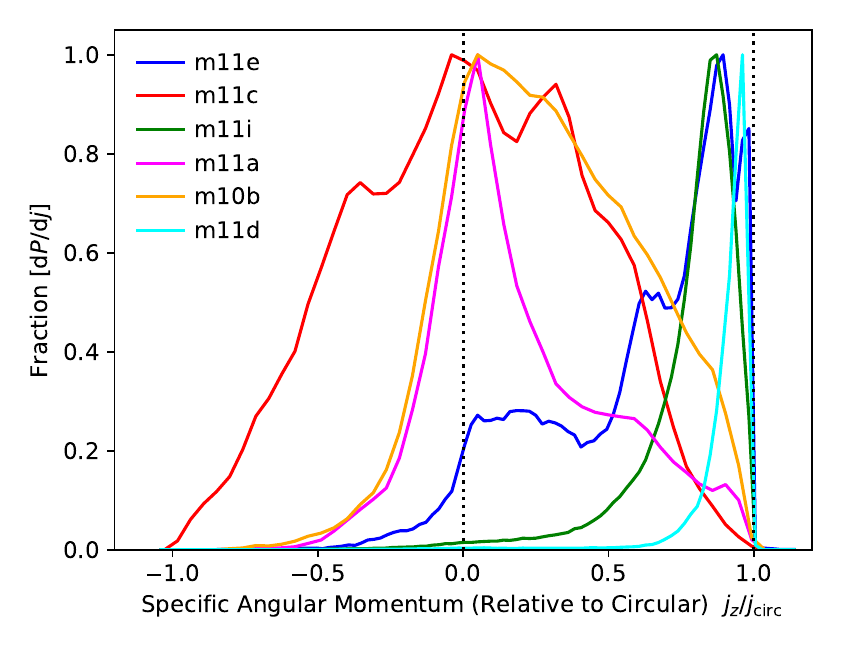}
    \caption{Distribution of specific angular momentum $j_z$ of HI gas relative to the angular momentum of a circular orbit, $j_{\rm{circ}}$, for several galaxies in our sample, with arbitrary normalization. The galaxies with the most obvious thin disks (e.g. m11d, m11i) are peaked near $j_z / j_{\rm{circ}} \sim 1$, while galaxies that are more pressure-supported than rotation supported (e.g. m11a, m11c, m10b) have wider distributions with a maximum near $j_z / j_{\rm{circ}} \sim 0$. }
    \label{fig:ang_mom}
\end{figure}

In simulations, we can determine whether a galaxy has a thin disk dominated by rotation by plotting the fraction of angular momentum of HI gas aligned with the axis of rotation (Figure \ref{fig:ang_mom}). Here, we clearly see that galaxies like m11d truly have thin disks. On the other extreme, the spherical, pressure supported dwarfs like m11c and m11a have peak at $j_z / j_{\rm{circ}} \sim 0$, where $j_z$ is the specific angular momentum, and $j_{\rm{circ}}$ is the angular momentum of a circular orbit. Galaxies like m11e or m11i, which could be mistaken for kinematically ordered, disky galaxies, fall somewhere in-between. 

\begin{figure*}
    \centering
    \includegraphics[width=0.9\textwidth]{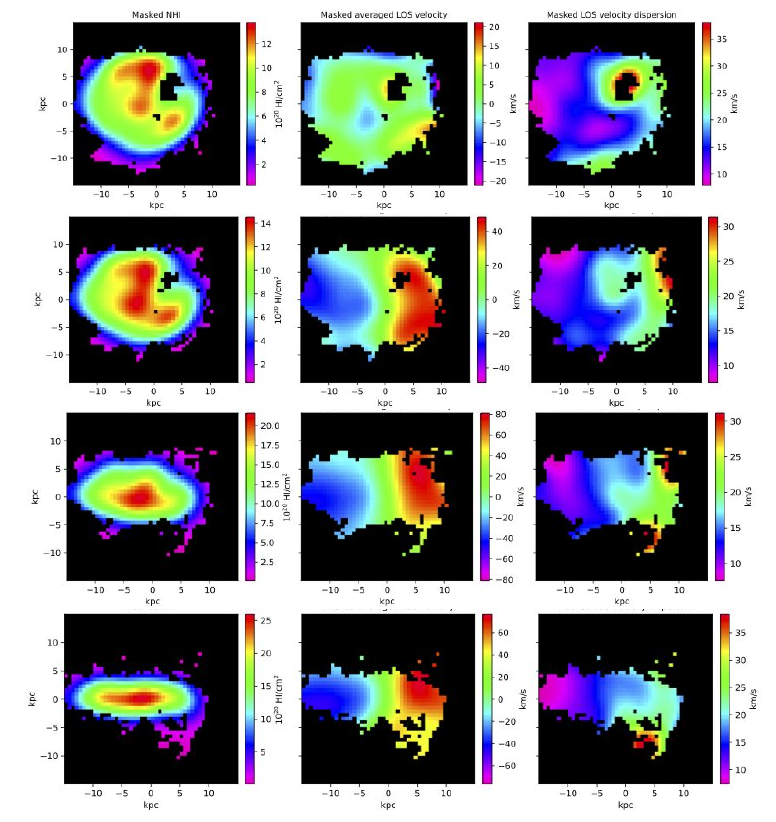}
    \caption{HI column density, average LOS velocity, and LOS velocity dispersion for FIRE galaxy m11i at $z = 0.02$, at varying viewing angles. This is a galaxy which has a turbulent disk with bursty star formation that varies in structure over time. The top row is face-on ($\phi=0$); $\phi$ increases by 30 degrees in each subsequent row, until m11i is projected to an edge-on view. Notable features that complicate RC analysis, such as the hole in the HI distribution, are hidden when the galaxy is viewed edge-on or nearly edge-on.}
    \label{fig:m11i_HI_angles}
\end{figure*}

However, there is no observable analog for this angular momentum metric. Instead, observers must use HI maps to identify the existence of a disk, but as we will demonstrate in this section, this data can be misleading. We use four visual indicators to determine if the galaxies in our sample are sufficiently disky for RC analysis: (1) A spiral, thin disk structure should be visible in the HI density map. As we will discuss in this section, the diskiness of a galaxy can be misconstrued depending on the viewing angle and resolution of observations. (2) The HI line-of-sight velocity map should show a velocity gradient centered at approximately 0 km/s. (3) The LOS velocity dispersion for an ordered, disky galaxy should be symmetric, with the highest velocity dispersion at the galactic center. (4) The ratio of the height of the galactic disk to its radius should be small. Large spiral galaxies with extended disks may have $h/R \sim 0.05-0.1$, but for dwarf galaxies, $h/R < 0.3$ is considered thin enough. 

These criteria are standard checks performed when selecting observed galaxies in a catalog for RC analysis; objects that do not meet the criteria are discarded \citep{Oh_2015}. Within our sample, we find that some of these criteria fail to rule out galaxies that do not have ordered disks, especially if only one or two out of all four criteria are used. In general, we find that the HI LOS velocity dispersion map and estimates of disk height-to-radius ratio are more faithful predictors of diskiness than the HI density map and average HI LOS velocity map, both of which are strongly dependent on viewing angle. 

The kinematic and dynamic complexity of some of the galaxies in our sample is not visually obvious from HI density and velocity moments alone. The viewing angle of the galaxy can conceal notable features in the HI distribution: for instance, in galaxy m11i at $z = 0.02$, stellar feedback blows HI gas out of a portion of the disk, leading to a large patch where the HI density falls below the simulated detection threshold of $5\times 10^{19} \rm{ HI cm}^{-2}$; we use this column density cutoff in order to consistently compare our mock HI maps to those from the Westerbork observations of neutral Hydrogen in Irregular and SPiral galaxies (WHISP) survey \citep{WHISP}. This hole is visually obvious when the galaxy is viewed face-on, but not from an edge-on view (Figure \ref{fig:m11i_HI_angles}). 

From the edge-on viewing angle, m11i appears to have more clear evidence of a thin disk at $z = 0.02$ than at $z = 0.04$ or $z = 0.0$, if only the HI column density and average LOS velocity are considered (Figure \ref{fig:m11i_HI}). The average LOS velocity for m11i at $z = 0.02$ shows a gradient consistent with rotation when viewed edge-on, albeit a somewhat lopsided one. However, the RC reconstructions for m11i are actually worse at $z = 0.02$ than in the other two shown (Figure \ref{fig:m11i_rc}), particularly in the inner few kpc. The galaxy height-to-radius ratio at $z = 0.0$ is approximately $0.2$, which, while not as thin as the more massive, disky m11s, is well within the bounds of what is considered disky for a dwarf galaxy.  It is not until the LOS velocity dispersion is considered that some of the complicating factors for m11i become apparent even when viewed perfectly edge-on: the velocity dispersion is not symmetric, and it peaks at the top edge of the disk rather than in the center. However, it should be noted that it is rare for dwarf galaxies to have perfectly symmetric LOS velocity dispersion, and that such galaxies are often included in catalogs for RC analysis \citep{HI_dwarf}.

\begin{table*}
    \hspace*{-1cm} 
    \centering
    \begin{tabular}{|c|c|c|c|c|c|} 
    \hline
           Galaxy name & $M_{\rm{HI}} (M_{\odot})$ & Distance (Mpc) & $R_{\rm{HI}}$ (kpc) & RC Behavior & Comparable FIRE galaxy \\ \hline 
          UGC 05721 & $5.61 \times 10^{8}$ & $6.18$ & $6.74$ & Cuspy inner region & FIRE-3 m11i \\ \hline
          UGC 08837 & $3.20 \times 10^{8}$ & $7.21$ & $4.75$ & linearly-rising core & FIRE-3 m11c \\ \hline
          IC 2574 & $1.036\times 10^{9}$ & $3.91$& $10.81$ & linearly-rising core & FIRE-2 m11e \\ \hline
    \end{tabular}
    \caption{A summary of the HI mass, distance, HI radius (defined as the point where HI surface density reaches $1$ $M_{\odot}\rm{pc}^{-2}$), and rotation curve behavior for the Spitzer Photometry \& Accurate Rotation Curves (SPARC) catalog dwarf galaxies used in comparisons to simulated FIRE galaxies in this section. We select the observational analogs to have similar rotation curve profiles and magnitudes to the RC reconstructions, as well as comparable HI masses to the respective FIRE galaxies. Values for HI mass, HI radius, and distance, as well as the measured rotation curves, are taken from the SPARC dataset \citep{2016AJ....152..157L}.}
    \label{tab:obs_gals}
\end{table*}

\subsubsection{A case study: ``contamination" with a sub-sample of FIRE-3 dwarfs}

One can imagine that galaxies like m11i could populate observational galaxy catalogs: features due to non-equilibrium phenomena may not be readily apparent from observations. It is difficult to estimate a contamination rate from our limited sample, especially since we have not attempted to replicate any galaxy catalog in particular here. However, as an illustrative example, we consider a set of mock HI surface density, LOS velocity, and velocity dispersion maps generated from two of the galaxies in our sample, m11i and m11d, at a uniformly-sampled range of viewing angles. We select these two FIRE-3 galaxies because m11i characterizes dwarf galaxies disrupted by bursty star formation and non-equilibrium effects, while m11d is more ordered and consistently produces faithful RC reconstructions at late times. In order to make the images more comparable with observational data from the WHISP survey, we pixelize the maps and apply a Gaussian filter with $\sigma$ equal to the beam size in kpc. Using WHISP's distance of $7.2$ Mpc for UGC 5721 measured with 60 arcsecond resolution, the beam size is $2.1$ kpc. 

We then apply the visual criteria described in the previous subsection in order to determine whether the HI data for the galaxy at the given viewing angle indicates that it is an acceptable candidate for RC analysis. Specifically, the HI surface density map must have a visible disk, a clear velocity gradient, and symmetry in the velocity dispersion. Large patches of missing HI or obvious signatures of large inflows and outflows are also considered disqualifying. We then calculate the ``rejection rate" of HI images for the galaxy at each of 5 late-time snapshots (ranging from $z = 0.04$ to $z = 0.0$). The results of this exercise are listed in Table \ref{tab:fail_rates}. We are fairly stringent in applying these criteria, and as such, the results of this exercise should be considered a conservative estimate. For instance, we reject the image of m11i shown in the middle panel of Figure \ref{fig:m11i_compare} due to asymmetries in the velocity dispersion, despite it having a similar degree of velocity dispersion asymmetry to UGC 5721, which is canonically included in most studies of the diversity problem.

\begin{table}
    \hspace*{-1cm} 
    \centering
    \begin{tabular}{|c|c|} 
    \hline
           Redshift & \% of viewing angles rejected for RC analysis  \\ \hline \hline
          \multicolumn{2}{|c|}{\bf{Galaxy m11i}} \\ \hline \hline
          $0.04$ & $27\%$ \\ \hline
          $0.03$ & $67\%$ \\ \hline
          $0.02$ & $83\%$ \\ \hline
          $0.01$ & $63\%$ \\ \hline
          $0.0$ & $83\%$ \\ \hline
          \hline \hline
          \multicolumn{2}{|c|}{\bf{Galaxy m11d}} \\ \hline \hline
          $0.04$ & $31\%$ \\ \hline
          $0.03$ & $13\%$ \\ \hline
          $0.02$ & $14\%$ \\ \hline
          $0.01$ & $21\%$ \\ \hline
          $0.0$ & $27\%$ \\ \hline
    \end{tabular}
    \caption{The percentage of mock HI images generated at uniformly sampled viewing angles from two FIRE-3 galaxies, m11i and m11d, that fail to meet the criteria established in the previous subsection to determine if the galaxy is sufficient disky, rotation-supported, and ordered to be used in RC analysis. We generate the images at five different late-time snapshots, spanning redshifts $z=0.04$ to $z=0.0$.}
    \label{tab:fail_rates}
\end{table}

Bearing in mind that the results of this exercise would provide a ``worst-case scenario" in terms of the fraction of HI images deemed acceptable for inclusion in RC analysis, we find that, as expected, m11i has higher rejection rates than m11d for nearly all snapshots. The rejection rates for m11i increase following the burst of late-time star formation after $z = 0.04$. However, upon closer examination, some of the pitfalls of relying on HI data alone to make judgments about fitness for RC analysis become evident. For example, the galaxy m11i snapshots at $z = 0.02$ and $z = 0.0$ have the same rejection rate (a relatively high 83\%), but, as shown in Figure \ref{fig:m11i_rc}, the RC reconstructions at $z=0.0$ more accurately trace the true $V_c$ than the RC reconstructions at $z=0.02$.  Moreover, we note that the rejection rate for galaxy m11d is not $0\%$ at any snapshot, despite having very faithful RC reconstructions. In other words, if we apply the same set of stringent selection criteria to a galaxy for which we know that rotation curves accurately trace the mass profile, there will still be viewing angles at certain times that show undesirable features. In m11d, the rejections mostly occur for viewing angles where one of the spiral arms covers the rest of the galaxy, giving the appearance of asymmetry in the HI velocity dispersion map. 

In many cases, whether the galaxy appears suitable for inclusion in RC analysis at a given viewing angle is debatable; virtually no galaxy (especially a dwarf) will meet all of the visual HI criteria, especially if not viewed directly edge-on. In these ambiguous cases, the inclusion of the galaxy in a catalog with rotation curves may rely upon a judgment call. As we will show in the following subsection, the inclusion of such galaxies could give rise to the appearance of more diversity in a set of rotation curves than truly exists in the underlying mass profiles. In light of this result, it would be desirable to develop more quantitative, robust metrics to determine which galaxies should be included in RC analysis. 

\vspace{1cm}

\subsection{Implications for the diversity problem}
\label{section:div_prob}
As demonstrated in the previous section, it is feasible that galaxies that are poor candidates for RC analysis can appear to meet the criteria for inclusion in galaxy catalogs. The inclusion of these galaxies can give rise to the appearance of ``artificial" rotation curve diversity, in which the spread of rotation curves from a galaxy catalog does not reflect true diversity of the underlying DM profiles, but rather, that there are galaxies for which inherently unmeasurable features are important. In this section, we compare our sample of simulated galaxies to several observed galaxies from the SPARC catalog \citep{2016AJ....152..157L}. A summary of the SPARC galaxies used and their HI masses, distances, and rotation curve behaviors is presented in Table \ref{tab:obs_gals}. 


\subsubsection{Artificial rotation curve diversity within FIRE dwarfs}

\begin{figure}
    \centering
    \includegraphics[width=\columnwidth]{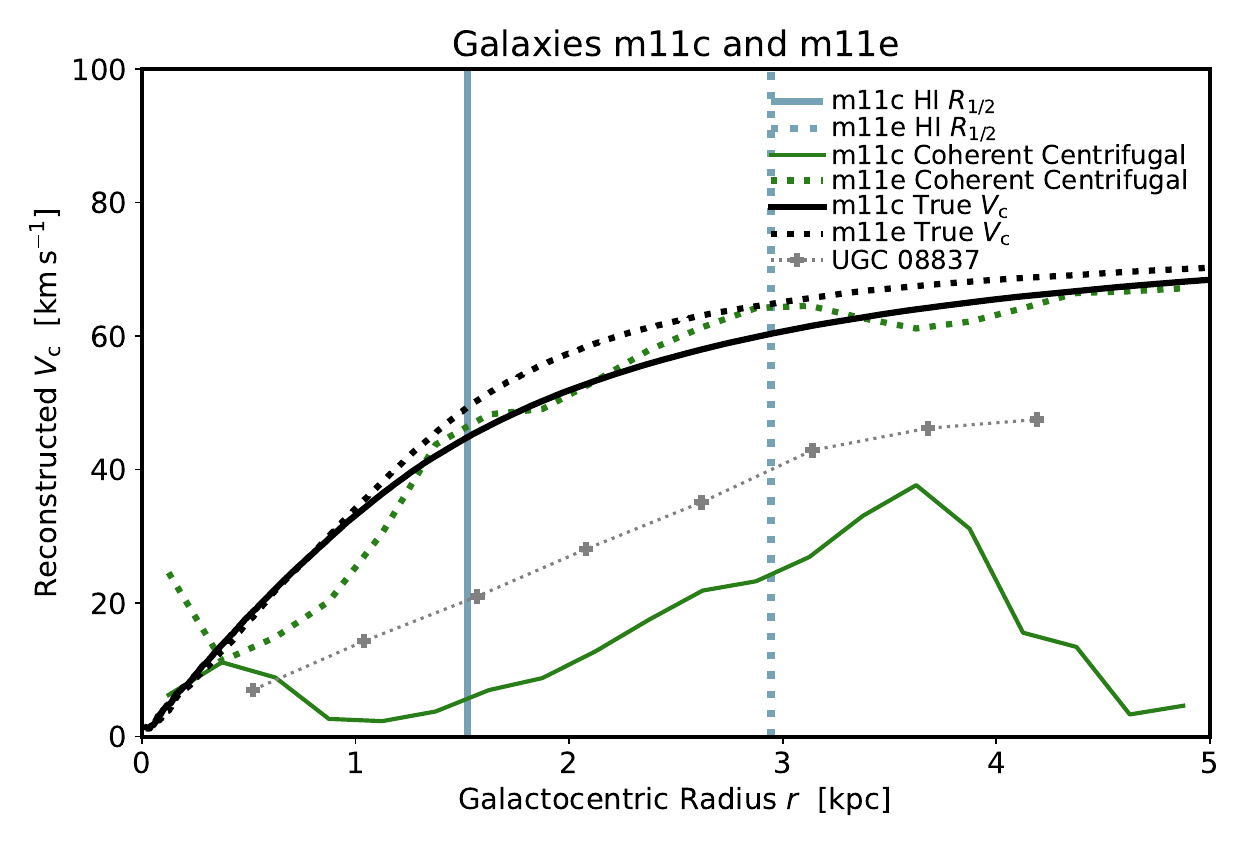}
    \caption{The true rotation curves and coherent centrifugal RC reconstructions for FIRE-3 galaxies m11c and m11e, along with the rotation curve measured for UGC 08837 \citep{2016AJ....152..157L}. While the two galaxies have fairly similar true $V_c$ curves, especially in the inner 1.5 kpc, the RC reconstructions wildly differ from each other. This difference is attributable to the difference in the astrophysical phenomena that shape each galaxy: m11c is dominated by magnetic fields, while m11e is rotationally-supported and subject to small tidal interactions with a nearby galaxy with which it will merge. FIRE-3 m11c shows a similar linearly-rising extended core to UGC 08837.}
    \label{fig:m11c_m11e_rc}
\end{figure}

While our sample of FIRE dwarf galaxies is too small to conduct a meaningful statistical analysis of variation in rotation curves, we are able to find cases where very different rotation curves are inferred from different galaxies with similar mass profiles, as is shown in Figure \ref{fig:m11c_m11e_rc}. FIRE-3 galaxies m11c and m11e have comparable true $V_c$ profiles: both show a slow rise in the central few kpc and level off around $65$ km/s. However, the coherent centrifugal rotation curve reconstructions (which depend only on the $\phi$ component of the centrifugal acceleration) for the two galaxies show strikingly different behavior. While the RC reconstruction for m11e maps the true $V_c$ reasonably well, the reconstruction for m11c considerably underestimates the true $V_c$ and rises linearly out to 4 kpc. We also show the measured rotation curve for galaxy UGC 08837, which has a linearly rising slope that extends to roughly the same radius as the coherent centrifugal RC reconstruction for m11c. 

\begin{figure}
    \centering
    \includegraphics[width=\columnwidth]{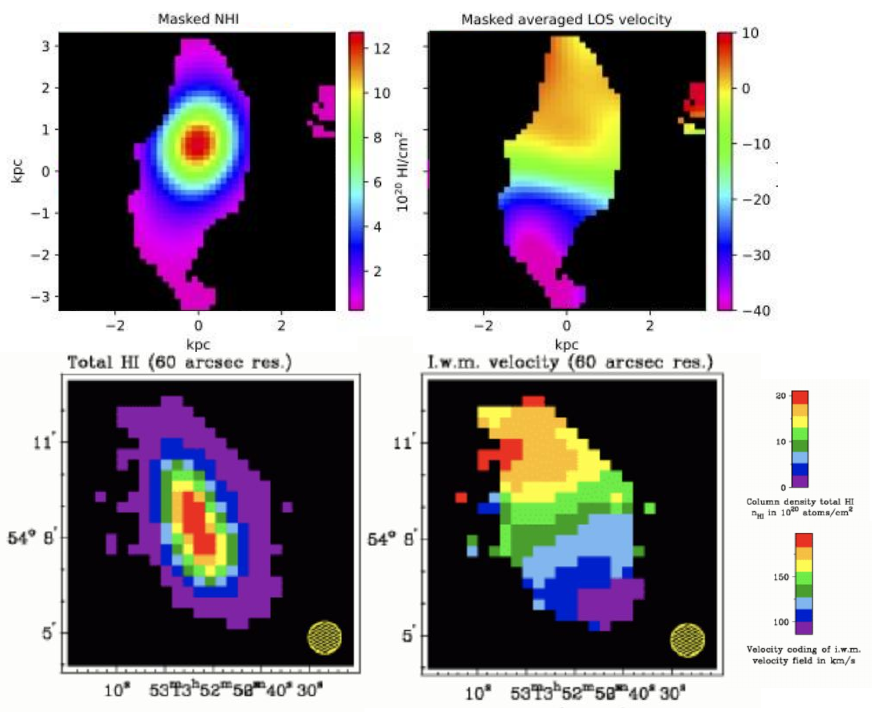}
    \caption{Top: HI column density and average LOS velocity for FIRE-3 m11c, pixelated and with a Gaussian blur applied to replicate observations. Bottom: HI column density and average LOS velocity for UGC 08837 from the WHISP survey \citep{WHISP}. Although FIRE-3 m11c does not form a disk and has little rotational support, its HI maps at this viewing angle show visual similarities to UGC 08837: it is roughly the same size and has similar concentric, elliptical contours in HI density. Additionally, both m11c and UG 08837 have asymmetric velocity shear. It is therefore plausible that a galaxy like m11c could be mistaken as disky and rotationally-supported.}
    \label{fig:m11c_ugc8837_hi}
\end{figure}

As discussed in section \ref{sec:cat_fail}, there are clear astrophysical explanations for the behaviors of these RC reconstructions. FIRE-3 galaxy m11e experiences a merger with another dwarf galaxy at late times; at the snapshot shown, the tidal effects of the merger are relevant in the outer regions of the galaxy, but the HI disk is still supported by rotation. In contrast, m11c is dominated by strong magnetic fields, and the HI exhibits little rotation. Since the coherent centrifugal reconstruction depends only on rotation, this RC reconstruction grossly underestimates the true $V_c$ of m11c. Thus, if the coherent centrifugal RC reconstructions for m11c and m11e are taken at face value as tracers of the galaxies' respective mass profiles, one would reach an incorrect conclusion about the amount of variation present in this sample. Instead of inferring the existence of two galaxies with rather similar mass profiles, there would appear to be one cored dwarf galaxy (m11e), and one much smaller dwarf with an extended, linearly-rising core (m11c). 

Moreover, the impact of the magnetic fields on the morphology and kinematics of m11c are not readily apparent from its HI maps. In Figure \ref{fig:m11c_ugc8837_hi}, we show the HI column density and average LOS velocity maps for m11c alongside the maps of UGC 08837 from the WHISP survey \citep{WHISP}. We select UGC 08837 for comparison because it is similar in size and HI mass to m11c, and its rotation curve is qualitatively similar to the reconstruction shown for m11c. Both m11c and UGC 0887 appear to have have concentric, elliptical density contours, as well as a clear (albeit asymmetric) velocity shear. As discussed in the previous subsection, these HI features are indicators of diskiness, and thus fitness for RC analysis. However, as shown in Figure \ref{fig:ang_mom}, m11c does not form a disk; its apparent velocity gradient is a result of viewing angle and the resolution and Gaussian blur used to generate HI maps that are comparable to real observations. 

The purpose of this comparison is not to state that UGC 08837 has similar strong magnetic fields to m11c, or even that it is a poor candidate for RC analysis, but simply to highlight the visual similarities between UGC 08837 and m11c. If UGC 08837 is included in RC analysis for SPARC-like catalogs, then it is plausible that a galaxy like m11c could be as well. Because the viewing angle and asymmetric velocity shear of m11c at $60$ arcsec resolution conceal features that complicate the interpretation of its rotation curve, the inclusion of such a galaxy in a rotation curve dataset would therefore lead to the appearance of ``artificial" diversity, with a RC with a linearly-rising slope inferred from a galaxy which has a more standard cored profile in actuality.

\subsubsection{Comparisons to observed diversity problem outlier galaxies}

\cite{Oman_2015} presents examples of observed objects that cannot be reproduced by CDM-only or hydrodynamic simulations with CDM, in particular, galaxies with extended cores or central profiles steeper than the NFW profile. In this section, we compare two of the observational outliers discussed in \cite{Oman_2015}, IC 2574 and UGC 5721, to simulated galaxies in our sample that show similar features in their reconstructed RCs. IC 2574 and UGC 5721 are frequently referenced in the literature as examples of the two extremes of the diversity problem: IC 2574 has a slowly-rising rotation curve with a roughly linear slope out to 12 kpc, while UGC 5721 has a very steep rise in circular velocity within the inner few kpc.

\begin{figure}[]
    \centering  
    \includegraphics[width=\columnwidth]{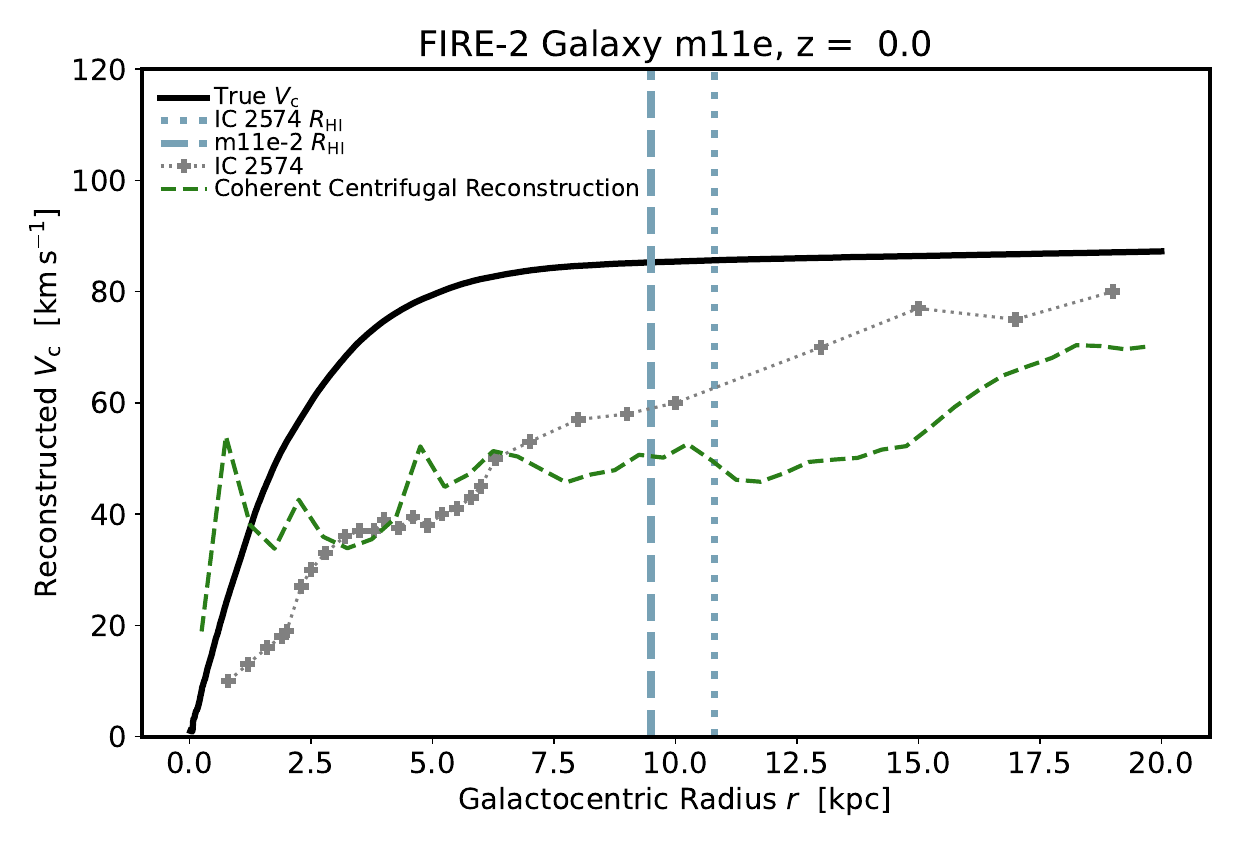}
    \caption{Rotation curve reconstruction for FIRE-2 m11e compared to the true $V_c$ defined by equation \ref{eqn:vc}; this galaxy is undergoing a merger. Using only the coherent centrifugal velocity to reconstruct the rotation curve for this galaxy, we see a linearly-rising slope, similar to the observed rotation curve of galaxy IC 2574, a galaxy that is commonly cited as one extreme of the rotation curve diversity problem \citep{Oman_2015}.}
    \label{fig:rc_IC 2574}
\end{figure}

Within our sample of simulated galaxies, we find many examples of galaxies with RC reconstructions that underestimate the true circular velocity, some of which give the appearance of more extended cores than the galaxy truly has. In galaxies with sufficient HI, this is usually due to dropped terms in the gravitational acceleration pertaining to non-kinetic and non-thermal stress (such as the presence of magnetic fields), or non-circular motions in the disk. In Figure \ref{fig:rc_IC 2574} we plot the RC of IC 2574 against the RC reconstructed from the perfect centrifugal moments of FIRE-2 galaxy m11e. We choose this FIRE-2 galaxy for comparison due to its RC reconstruction's similarity in linear slope and overall magnitude to IC 2574. 

\cite{2020MNRAS.493.2618S} conduct a detailed analysis of the HI distribution and kinematics of IC 2574 and its nearby HI complex, HIJASS J1021+68, and find that this galaxy is highly asymmetric in both its HI density distribution and its first and second kinematic moments. Moreover, the HI distribution in IC 2574 is notable for containing multiple HI holes \citep{2007ApJ...661..102W}. FIRE-2 m11e is characterized by many of these same features: there is a large hole in the HI, and the first and second kinematic moments are highly asymmetric. This galaxy is also the result of a recent merger, and from certain viewing angles, the signature tidal tail is easily visible. Other viewing angles, however, completely obscure the tidal tail, and give the appearance of an irregularly-shaped HI region with a nearby companion. The hidden tidal tail in m11e-2 bears resemblance to HIJASS J1021+68; however, whether HIJASS J1021+68 is a tidal feature or merger remnant like the analogous structure in m11e-2 is a matter of debate. While \cite{2020MNRAS.493.2618S} argue that HIJASS J1021+68  is unlikely to be a tidal dwarf, as IC 2574 does not appear to have a tidal tail, \cite{Boyce_2001} suggest that the HI complex may be the remnant of a tidal interaction with IC 2574. Regardless, IC 2574 and HIJASS J1021+68 are clearly a system with a significant degree of non-equilibrium behavior. That m11e-2, a simulated galaxy with similar morphological features, has a strikingly similar reconstructed RC suggests that these non-equilibrium effects are likely essential to characterizing the circular velocity and underlying dark matter profile for IC 2574. These features may render IC 2574 a poor choice of galaxy to characterize the diversity problem, and possibly provide evidence for a merger. 

RC reconstructions that are ``cuspier" than the true profile are more rare in our sample. In most of our sample galaxies where we see these kinds of RC reconstructions, this error appears to be a fluctuation-- in the next snapshot, the same reconstruction method may under-predict the true circular velocity, or be closer to the true value (Figure \ref{fig:m11i_rc}). It also only appears in the less-ordered galaxies in our sample, those with significant inflows, outflows, torques, or no true gaseous disk. This suggests that the apparent mass excess in the inner region of these galaxies is due to time-dependent effects that inherently cannot be detected by observations.

Galaxy m11i has reconstructed RCs at $z = 0.02$ that show this excess cuspiness in the inner two kpcs. As described in the previous section, m11i features a large area in the HI disk where stellar feedback has blown out gas, leading to the appearance of a ``hole" (Figure \ref{fig:m11i_HI_angles}). The HI gas around the hole gets a velocity kick, which makes m11i look like it is rotating faster in the inner region than it actually is; this ultimately leads to the ``cuspy" RC reconstructions in the middle panel of Figure \ref{fig:m11i_rc}. We compare the RC reconstructions to the RC measured for UGC 5721, one of the galaxies described in \cite{Oman_2015} as more cuspy than an NFW profile for a halo of its virial mass. The RCs for m11i that include information about radial and azimuthal velocity in addition to the purely rotational velocity (shown in red and cyan) have a sharp spike that quickly falls off and flattens throughout the rest of the HI disk, which is similar to the RC for UGC 5721,

While the star formation history of UGC 5721 is an area of ongoing research, there is evidence that suggests burstiness. One criterion commonly used in the literature to classify galaxies as starbursts is that the ratio of the Infrared Astronomical Satellite (IRAS) 60 and 100 $\rm{\mu}m$ flux densities (in Jy) exceeds $0.5$ \citep{IRAS1989MNRAS.238..523R}; for UGC 5721, $f_{60}/f_{100} = 0.53$ using the IRAS Point Source Catalog (PSC) measurements, and $f_{60}/f_{100} = 0.51$ using the IRAS Faint Source Catalog (FSC) measurements \citep{1990IRASF.C......0M, 1994yCat.2125....0J}. In addition, \citet{Lelli_2014} identify UGC 5721 as a potential starbursting dwarf based on its dynamical and structural properties. Thus, it is feasible that bursty star formation could influence the rotation curve measured for this galaxy.

\begin{figure*}[]
    \centering
    \includegraphics[width=\textwidth]{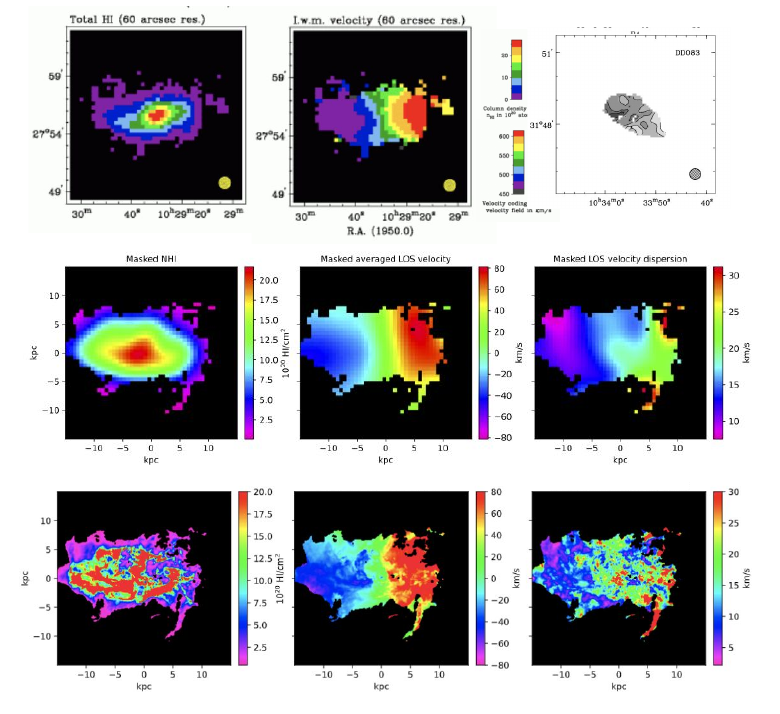}
    \caption{Top: HI column density and average LOS velocity for UGC 5721, from the WHISP survey \citep{WHISP}; LOS velocity dispersion for UGC 5721 from \cite{HI_dwarf}. Middle: HI column density and average LOS velocity 
    FIRE galaxy m11i at $z = 0.02$, pixelated and with a Gaussian blur applied. Bottom: HI column density and average LOS velocity for FIRE galaxy m11i with the filters removed. The mock observations of m11i appear regular and comparable to UGC 5721, concealing evidence of the compounding factors (i.e. bursty star formation) that cause an anomalous cusp in the RC reconstruction for m11i. The inclusion of a galaxy like m11i in an observational sample for RC analysis based on its regular appearance would therefore create ``artificial" diversity in the spread of RCs.}
    \label{fig:m11i_compare}
\end{figure*}

In Figure \ref{fig:m11i_compare}, we show the HI column density and LOS velocity for UGC 5721, and compare it to the HI maps for m11i at $z = 0.02$ at a comparable viewing angle (not exactly edge-on and slightly tilted, and with the hole not visible). In the middle panel, we show the HI maps for m11i at a low resolution and with a gaussian blur applied to make it more comparable to observed HI data. In the bottom panel, we show the same HI maps for m11i, but with the filters and pixelization removed. There are significant visual similarities between the processed image of m11i and UGC 5721: there is a gradient in the LOS velocity, but it is slightly asymmetric. The HI LOS velocity dispersion, as shown in \cite{HI_dwarf}, shows a much more significant asymmetry. Although m11i looks regular in the mock observations, it has underlying non-equilibrium dynamics that cause it to have an anomalously-cusped reconstructed RC. Such systems would be included in RC samples and introduce ``false" diversity.

Neither IC 2574 nor UGC 5721 are perfectly ``clean" galaxies in their HI distributions and kinematic moments-- but neither are the simulated dwarfs within our sample. While this messiness is to be expected in the dwarf galaxy regime, it is important to use caution when interpreting results from these galaxies, especially for derived measurements like rotation curves. We have shown that a variety of dynamical phenomena can reproduce the qualitative behaviors seen in the RCs for the most extreme galaxies cited as examples of the diversity problem. While some of this diversity may be real, at least some portion of it may be due to unmeasurable components of the galaxy kinematics and dynamics that are not accounted for in RC analysis.

\section{Summary and Conclusion} 
\label{section:conclusion}

In this work, we have evaluated a sample of FIRE dwarf galaxies in order to determine what kinematic and dynamical information is necessary to recover the underlying circular velocity profile that maps the enclosed mass. We have assumed perfect knowledge of the gravitational acceleration, time-dependent terms, and stress tensor components-- this study provides a best-case, idealized scenario for rotation curve analysis where there are no observational uncertainties. Even after assuming perfect data, we find that attempts to recover the rotation curve may deviate significantly from the true circular velocity. In dwarf galaxies where a thin, ordered disk forms, these deviations are relatively minor, within approximately 10\% error. However, most of the dwarfs in our sample are subject to dynamical effects that lead to more complex systems that the ideal thin, rotation-dominated disk. In these cases, we find several common failure modes. Dwarfs with particularly small, puffy disks may lack enough HI for faithful RC reconstructions. Other dwarfs are dominated by  components that are very difficult to detect through observations, such as strong magnetic fields; this usually leads to a significant underestimate of the true circular velocity. Finally, some dwarfs have significant time-dependent effects due to phenomena like bursty star formation or galactic mergers. Depending on the viewing angle and resolution of the HI observations, signatures of these phenomena may not be clearly visible. 

We have shown that the effect of these dynamic, non-equilibrium, and often unobservable astrophysical phenomena can be to change the shapes of the apparent rotation curves in a number of ways: often, the rotation curve reconstructions show much more variation in overall shape and features than the profiles for true circular velocity. We consider several case study comparisons to \cite{Oman_2015} extrema dwarf galaxies for opposite ends of diversity problem. The observed dwarfs IC 2574 and UGC 5721 are considered canonical examples of diversity problem galaxies, and both have been categorized as sufficiently HI-rich and disky for inclusion in standard RC analyses. Within a relatively small sample of simulated galaxies, we find instances of FIRE dwarfs where the reconstructed RCs replicate features of the rotation curves for these two galaxies: an extended core for IC 2574 (in analogy to m11e-2), and an overly-cuspy inner profile for UGC 5721 (m11i). It is important to note that these RC features represent the extremes of the diversity problem, and occur relatively rarely within observational galaxy catalogs (roughly $\sim 1$ in $100$ galaxies). That we are able to reproduce these features within a sample of $16$ FIRE galaxies suggests that astrophysical phenomena add an intrinsic diversity to the shapes of RCs, and therefore directly bear on the diversity problem. It is feasible that some of the diversity problem outlier galaxies may be poor candidates for RC analysis, and their inclusion in rotation curve datasets may create the appearance of more RC diversity than truly exists within a catalog simply because what is measured does not accurately trace the enclosed mass of the galaxy. 

Our results beg the question: how often can we expect the contamination of galaxy catalogs by non-equilibrium, asymmetric dwarfs to occur? As we have shown, it is difficult to know purely from observational features whether a galaxy meets the assumptions for RC analysis. A robust answer to these questions would require a uniformly volume-limited sample of dwarf galaxies in real HI data cube space, with a specific survey's selection function applied. Then, forward-modeled observables from these simulations, such as the luminosity and width of the unresolved HI spectrum, velocity shear, and velocity dispersion, can be compared to observed HI data cubes. The advantage of this comparison is that it would not rely on parametrizations, such as tilted-ring fitting in rotation curve analysis, and would make fewer assumptions overall.  In light of the results of this work, we will explore the efficacy of such alternate methods to rotation curve analysis in future work. 

It is critical to determine the extent to which rotation curve diversity is attributable to intrinsic variation in halo mass profiles as opposed to galaxy-scale astrophysics. Such work will have important and dramatic consequences for the wealth of alternative dark matter models that have been posed as conceivable explanations for the diversity problem: the diversity problem only poses a challenge to $\Lambda$CDM if it reflects variation in mass profiles. To this end, the diversity problem should not be dismissed, but rather, re-characterized in terms of metrics that are less sensitive to non-equilibrium and complex galactic dynamics. 

\section*{Acknowledgments}
\label{sec:acknowledgements}
Support for ISS and PFH was provided by NSF Research Grants 1911233, 20009234, 2108318, NSF
CAREER grant 1455342, NASA grants 80NSSC18K0562, and HST-AR15800. MBK acknowledges support from NSF CAREER award AST-1752913, NSF grants AST-1910346 and AST-2108962, NASA grant 80NSSC22K0827, and HST-AR-15809, HST-GO-15658, HST-GO-15901, HST-GO-15902, HST-AR-16159, HST-GO-16226, HST-GO-16686, HST-AR-17028, and HST-AR-17043 from the Space Telescope Science Institute, which is operated by AURA, Inc., under NASA contract NAS5-26555. JSB was supported by NSF grant AST-1910965 and NASA grant 80NSSC22K0827. CAFG was supported by NSF through grants AST-2108230, AST-2307327, and CAREER award AST-1652522; by NASA through grants 17-ATP17-0067 and 21-ATP21-0036; by STScI through grants HST-GO-16730.016-A and JWST-AR-03252.001-A; and by CXO through grant TM2-23005X. JM is funded by the Hirsch Foundation. FJM is funded by the National Science Foundation (NSF) Math and Physical Sciences (MPS) Award AST-2316748. Numerical calculations were run on the Caltech compute cluster ``Wheeler," allocation AST21010 supported by
the NSF and TACC, and NASA HEC SMD-16-7592. This research has made use of the VizieR catalogue access tool, CDS, Strasbourg, France \citep{10.26093/cds/vizier}. The original description 
of the VizieR service was published in \citet{vizier2000}.


\bibliographystyle{aasjournal}

\clearpage

\appendix

\subsection{Rotation Curve Details}
\label{section:app_RC}
In hydrodynamic simulations, the velocities of stars and gas particles are determined by the Vlaslov-Poisson equation: 

\begin{align}
\label{eqn:vpe.general} \frac{\partial f}{\partial t} + {{\bf v}} \cdot \nabla f + {\bf F} \cdot \frac{\partial f}{\partial {\bf p}} = \left( \frac{\partial f}{\partial t} \right)_{\rm collisions},
\end{align}
where $f$ is the distribution function for the particle species, $\mathbf{F}$ is the force acting on the particle, and $\mathbf{p}$ is the momentum. At every time step in the simulations, the GIZMO code solves the momentum equation: 
\begin{align}
-\nabla \Phi &\equiv {\bf a}_{\rm grav} \equiv \frac{1}{\rho} \left[  \frac{\partial (\rho {\bf v} )}{\partial t} + \nabla \cdot \boldsymbol{\Pi}^{\ast} - {\bf S}_{\rm ext} \right],
\label{eqn:momentum}
\end{align}
where $\Phi$ is the acceleration due to gravity, and ${\bf S}_{\rm ext}$ is the external source term due to radiation and cosmic rays, with ${\bf S}_{\rm ext} \equiv \rho\, {\bf a}_{\rm rad} + {\bf s}_{\rm cr} $, ${\bf a}_{\rm rad} \equiv  \frac{1}{c}\int\,\kappa_{\nu}{\bf F}_{\nu}\,d\nu $, and ${\bf s}_{\rm cr} \equiv -\frac{1}{c^{2}} \hat{\bf B}\, D_{t} F_{\rm cr,\,e}$.
$\boldsymbol{\Pi}^{\ast}$ is the stress tensor, which includes components from the gas kinematics, magnetic field, thermal pressure, viscosity, and cosmic rays: 
\begin{align}
\boldsymbol{\Pi}^{\ast} &\equiv \boldsymbol{\Pi}_{\rm kin} + \boldsymbol{\Pi}_{\rm mag} + \boldsymbol{\Pi}_{\rm therm} + \boldsymbol{\Pi}_{\rm visc} + \boldsymbol{\Pi}_{\rm cr},
\label{eqn:pi_comps}
\end{align}
with the kinetic stress defined as $\boldsymbol{\Pi}_{\rm kin} \equiv \rho {\bf v} {\bf v}$, thermal stress as $\boldsymbol{\Pi}_{\rm therm}  \equiv n k_{\rm B} T {\bf I} $, magnetic stress as $\boldsymbol{\Pi}_{\rm mag} \equiv \frac{{\bf B}\cdot{\bf B}}{8\pi} {\bf I} - \frac{{\bf B}{\bf B}}{4\pi} $, stress due to viscosity of the gas as $\boldsymbol{\Pi}_{\rm visc} \equiv \frac{\nu_{\rm visc}}{3}\,\left( 3 \hat{\bf B} \hat{\bf B} - {\bf I} \right) \left(3 \hat{\bf B} \hat{\bf B} - \bf{I} \right) : \left(\nabla {\bf v} \right) $, and stress due to cosmic rays as $\boldsymbol{\Pi}_{\rm cr} \equiv \int {\bf p}_{\rm cr}\,{\bf v}_{\rm cr}({\bf p}_{\rm cr}) f_{\rm cr}({\bf p}_{\rm cr}) \, d^{3}{\bf p}_{\rm cr}$. $\rho$ is the density, $\bf{I}$ is the identity tensor, $T$ is the gas temperature, and $\bf{B}$ is the magnetic field. 

We now consider several approximations to the fully general equation \ref{eqn:momentum}, which will be evaluated in turn through the subsequent sections. First, we average over the gravitational acceleration in an annulus in the plane perpendicular to the $z$-component of the angular momentum vector. In plots, we refer to this approximation as the full rotation curve reconstruction, assuming all components of the stress tensor, non-equilibrium terms, and external source terms are measurable. 

Since observations usually can only measure stellar and/or gas velocities at a single instance in time, the time-derivative term in equation \ref{eqn:momentum} is, effectively, unmeasurable. Thus, we drop this term, along with the the external source term; the latter is usually negligible compared to the components of the stress. We refer to the reconstructed RC with these terms dropped as the equilibrium reconstruction. 

Next, we consider the components of the stress tensor that cannot be measured from observations. These include the tensor components pertaining to the magnetic field, viscosity, and cosmic rays, which are often neglected in observational RC analysis. Dropping these terms, expanding the Laplacian in spherical coordinates, and taking only the radial component of the gravitational acceleration, we obtain 

\begin{align}
\langle {\bf a}_{\rm grav} \rangle &\underset{\mathrm{drop}}{\sim} \left\langle \frac{1}{\rho}  \nabla \cdot \left( \boldsymbol{\Pi}_{\rm kin} + \boldsymbol{\Pi}_{\rm therm} \right) \right\rangle 
= \langle \frac{1}{\rho} 
\left[ 
\frac{1}{r^{2}}\frac{\partial (r^{2} \rho v_{r}^{2})}{\partial r}
+ \frac{1}{r\,\sin{\theta}} \frac{\partial (\rho\,v_{\theta}\,v_{r}\,\sin{\theta})}{\partial \theta} 
\right. 
+ \left. \left. \frac{1}{r\,\sin{\theta}} \frac{\partial (\rho\,v_{\phi}\,v_{r})}{\partial \phi}
- \frac{\rho\,(v_{\theta}^{2} + v_{\phi}^{2})}{r}
+ \frac{\partial P}{\partial r}
\right] \right\rangle .
 \label{eqn:kin_therm}
\end{align} 

The contributions from these non-kinetic stress tensor terms almost always end up being relatively small. More significant deviations occur when we begin to make assumptions about the dominant components of the HI velocities. If we assuming the galaxy is midplane-symmetric ($\partial_{z} \rightarrow 0$) and azimuthally symmetric ($\partial_{\phi} \rightarrow 0$), then equation \ref{eqn:kin_therm} further reduces to
\begin{align}
\langle { a}_{\rm grav} \rangle &\underset{\mathrm{sym}}{\sim} 
 \left\langle \frac{1}{\rho} 
\left[ 
\frac{1}{r^{2}}\frac{\partial (r^{2} \rho v_{r}^{2})}{\partial r}
- \frac{\rho\,(v_{z}^{2} + v_{\phi}^{2})}{r}
+ \frac{\partial P}{\partial r}
\right]
 \right\rangle .
\end{align}
We refer to this RC reconstruction as the non-circular recovery reconstruction. 

Finally, we make the most stringent assumption: that the only measurable component of velocity is the centrifugal $v_{\phi}$, and that the LOS-dispersion components from thermal pressure cannot be recovered. Then, the acceleration due to gravity becomes 
\begin{align}
\langle { a}_{\rm grav} \rangle &\underset{\mathrm{sym}}{\sim} 
 \left\langle 
- \frac{ v_{\phi}^{2}}{r}
 \right\rangle ,
\end{align}
which we refer to as the coherent centrifugal reconstruction. 

\end{document}